\documentclass[12pt]{article}
\usepackage[pdftex]{graphicx} %%GRAPHICX
\usepackage{epsfig}
\usepackage{comment}
\usepackage{latexsym}
\usepackage{hyperref}
\usepackage{amsmath}
\usepackage{color}

\newcommand{\mysquare}[0]{\raise-.2ex\hbox{{\Large$\Box$}}}
\def\lsim{\mathrel{\rlap {\raise.5ex\hbox{$ < $}}
{\lower.5ex\hbox{$\sim$}}}}
\def\gsim{\mathrel{\rlap {\raise.5ex\hbox{$ > $}}
{\lower.5ex\hbox{$\sim$}}}} \topmargin -1.5cm \textheight=22.5cm \textwidth=16.5cm
\catcode`\@=11
\newcount\hour
\newcount\minute
\newtoks\amorpm
\hour=\time\divide\hour by60 \minute=\time{\multiply\hour by60
\global\advance\minute by-\hour}
\edef\standardtime{{\ifnum\hour<12 \global\amorpm={am}%
        \else\global\amorpm={pm}\advance\hour by-12 \fi
        \ifnum\hour=0 \hour=12 \fi
        \number\hour:\ifnum\minute<10 0\fi\number\minute\the\amorpm}}
\edef\militarytime{\number\hour:\ifnum\minute<10 0\fi\number\minute}
\def\draftlabel#1{{\@bsphack\if@filesw {\let\thepage\relax
   \xdef\@gtempa{\write\@auxout{\string
      \newlabel{#1}{{\@currentlabel}{\thepage}}}}}\@gtempa
   \if@nobreak \ifvmode\nobreak\fi\fi\fi\@esphack}
        \gdef\@eqnlabel{#1}}
\def\@eqnlabel{}
\def\@vacuum{}
\def\draftmarginnote#1{\marginpar{\raggedright\scriptsize\tt#1}}
\def\draft{\oddsidemargin -.2truein
        \def\@oddfoot{\sl preliminary draft \hfil
        \rm\thepage\hfil\sl\today\quad\militarytime}
        \let\@evenfoot\@oddfoot \overfullrule 3pt
        \let\label=\draftlabel
        \let\marginnote=\draftmarginnote
   \def\@eqnnum{(\theequation)\rlap{\k

 ern\marginparsep\tt\@eqnlabel}%
\global\let\@eqnlabel\@vacuum}  }
\newcommand{\be}[0]{\begin{equation}}
\newcommand{\ee}[0]{\end{equation}}
\newcommand{\ba}[0]{\begin{eqnarray}}
\newcommand{\ea}[0]{\end{eqnarray}}
\newcommand{\dis}{\displaystyle}

\def\bs{\begin{subequations}}
\def\es{\end{subequations}}

\def\thebibliography#1{%
\vskip 0.5cm \centerline{\bf \Large References}
\list{%
[\arabic{enumi}]}{\settowidth\labelwidth{[#1]} \leftmargin\labelwidth
\advance\leftmargin\labelsep
\usecounter{enumi}}
\def\newblock{\hskip .11em plus .33em minus .07em}
\sloppy\clubpenalty4000\widowpenalty4000 \sfcode`\.=1000\relax}

\renewcommand{\theequation}{\arabic{section}.\arabic{equation}}

\renewcommand{\section}{\setcounter{equation}{0}\@startsection
{section}{1}{0mm}{-\baselineskip}{0.5\baselineskip} {\normalfont\Large\bfseries}}

\renewcommand{\subsection}{\@startsection
{subsection}{2}{0mm}{-\baselineskip}{0.5\baselineskip} {\normalfont\large\bfseries}}

\renewcommand{\subsubsection}{\@startsection
{subsubsection}{3}{0mm}{-\baselineskip}{0.5\baselineskip}
{\normalfont\normalsize\slshape}}

\usepackage{amssymb}
\usepackage{amsthm}
\usepackage{amsmath}
\usepackage{amssymb,amsfonts}
\usepackage{graphicx}
\usepackage{cite}

\newcommand{\ie}{{\em i.e. }}

\newcommand{\Z}{\mathbb{Z}}
\newcommand{\sign}{\mbox{sign}}

\renewcommand{\O}{{\cal O}}
\renewcommand{\Re}{{\rm Re}}
\renewcommand{\Im}{{\rm Im}}
\newcommand{\abs}{|}
\newcommand{\where}{\mbox{where}}

\renewcommand{\and}{\mbox{and}}
\newcommand{\esp}{\!\!\!\phantom{\Bigg\abs}}
\newcommand{\desp}{\!\!\! \phantom{\underset{\hat \abs}{\Big\abs}}}
\newcommand{\tesp}{\!\!\! \phantom{\underset{\Big\abs}{\big\abs}}}
\newcommand{\bbf}{\bf\boldmath}

\newcommand{\N}{{\cal N}}
\renewcommand{\S}{{\cal S}}

\begin{document}
%\verb|\usepackage{draftcopy}|\\
\begin{titlepage}
\begin{flushright}
LPTENS--11/20,
CPHT--RR046.0511,
June 2011
\vspace{-.0cm}
\end{flushright}
\begin{centering}
{\bf \Large Thermal Duality and Non-Singular Cosmology  \\ 
\vskip 0.2cm
in $d$--dimensional\ Superstrings$^*$ }

\vspace{5mm}

 {\bf Costas Kounnas$^{1}$, Herv\'e Partouche$^2$ and Nicolaos Toumbas$^3$}

\vspace{1mm}

$^1$ Laboratoire de Physique Th\'eorique,
Ecole Normale Sup\'erieure,$^\dag$ \\
24 rue Lhomond, F--75231 Paris cedex 05, France\\
{\em  Costas.Kounnas@lpt.ens.fr}

$^2$  {Centre de Physique Th\'eorique, Ecole Polytechnique,$^\ddag$\\
F--91128 Palaiseau cedex, France\\
{\em herve.partouche@cpht.polytechnique.fr}}

$^3$  Department of Physics, University of Cyprus,\\
Nicosia 1678, Cyprus.\\
{\em nick@ucy.ac.cy}

\end{centering}
\vspace{0.1cm}
$~$\\
\centerline{\bf\Large Abstract}\\
\vspace{-0.2cm}

%\begin{quote}
\noindent 
We are presenting the basic ingredients of a stringy mechanism able to resolve both the Hagedorn instabilities of finite temperature superstrings as well as the initial singularity of the induced cosmology in arbitrary dimensions. These are shown to be generic in a large class of $(4,0)$ type II superstring vacua, where non-trivial ``gravito-magnetic'' fluxes lift the Hagedorn instabilities of the thermal ensemble and the temperature duality symmetry is restored. This symmetry implies a universal maximal critical temperature. In all such models there are three characteristic regimes, each with a distinct effective field theory description: Two dual asymptotically cold regimes associated with the light thermal momentum and light thermal winding states, and the intermediate regime where additional massless thermal states appear. The partition function exhibits a conical structure as a function of the thermal modulus, irrespectively of the spacetime dimension. Thanks to asymptotic right-moving supersymmetry, the genus-1 partition function is well-approximated by that of massless thermal radiation in all of the three effective field theory regimes. The resulting time-evolution describes a bouncing cosmology connecting, via spacelike branes, a contracting  thermal ``winding"  Universe to an expanding thermal ``momentum"  Universe, free of any essential curvature singularities. The string coupling remains perturbative throughout the cosmological evolution. Bouncing cosmologies are presented for both zero and negative spatial curvature. 
%\end{quote}

\vspace{3pt} \vfill \hrule width 6.7cm \vskip.1mm{\small \small \small
  \noindent
   $^\dag$\ Unit{\'e} mixte  du CNRS et de l'Ecole Normale Sup{\'e}rieure associ\'ee \`a
l'Universit\'e Pierre et Marie Curie (Paris 6), UMR 8549.\\
$^\ddag$\ Unit{\'e} mixte du CNRS et de l'Ecole Polytechnique,
UMR 7644.}\\

\end{titlepage}
\newpage
\setcounter{footnote}{0}
\renewcommand{\thefootnote}{\arabic{footnote}}
 \setlength{\baselineskip}{.7cm} \setlength{\parskip}{.2cm}

\setcounter{section}{0}

%%%%%%%%%%%%%%%%%%%%%%%%%%%%%%%%%%%%
%%%%%%%%%%%%%%%%%%%%%%%%%%%%%%%%%%%%

\section{Introduction}

Observational evidence strongly supports that during an early 
cosmological era, the matter content of the Universe was in 
(near) thermal equilibrium, with very high temperature. If the
degrees of freedom are to be described by a set of local quantum fields, 
such a state results in a singular cosmology. Indeed, if we  
follow the cosmological evolution backward in time, using Einstein's 
gravity field equations, we are driven   
to the initial curvature singularity\cite{HE}. 
Even if a period of inflation preceded the high temperature phase,
it is found in typical field theory models that 
the cosmological evolution begins at a singularity.

In string theory we expect a drastically different picture to emerge since
new purely stringy degrees of freedom can dominate the high curvature 
and high temperature regimes, leading to phenomena 
that do not admit a conventional 
field theory description \cite{CosmoTopologyChange}, with Riemannian concepts breaking down. 
String oscillators and winding states become relevant around the Hagedorn 
temperature $T_H$ (which is of order the string scale $M_s$), before the onset of curvature singularities, and  
drive a phase transition towards a new stringy thermal vacuum 
\cite{CosmoPheno,GV,AW,RK,AKADK,BR,DLS,Chaud,DL,HotstringsCosmo}. 
The simplest way to isolate the relevant critical phenomena is
via the Euclidean description of the thermal system, where Euclidean
time is compactified on a circle with period
given by the inverse temperature \cite{AW,RK,AKADK}. At temperatures just above Hagedorn,
certain string states winding the Euclidean time circle become tachyonic. 
These instabilities can be lifted either by condensing the tachyons \cite{AW,AKADK}, or by
turning on special gravito-magnetic fluxes, which inject into the thermal vacuum
non-trivial winding and momentum charges, as in \cite{akpt,massivesusy,FKT,FKPT}. If a 
stable stringy phase gets realized, it could be that the back-reacted cosmological evolution is 
non-singular and the initial singularity is absent.  

A mechanism within which the Hagedorn instabilities of the string gas 
are resolved and the initial curvature singularity is bypassed 
was realized recently in a class of two-dimensional 
superstring cosmologies, the so called Hybrid cosmologies \cite{FKPT}. 
The scope of the present work is to show that the key ingredients 
of this mechanism are generic in a diverse class of higher 
dimensional superstring models as well. 
In all of these models, finite temperature is introduced 
along with non-trivial gravito-magnetic fluxes \cite{akpt,massivesusy,FKT,FKPT}, 
which lead to a restoration of the thermal duality symmetry 
of the partition function:  $Z(\beta / \beta_c)= Z(\beta_c/\beta)$. 
Here $\beta$ denotes the period of the Euclidean time cycle, 
attaining a critical value $\beta_c$ at the self-dual point. 
At this critical point additional massless thermal states appear, 
enhancing the local Euclidean gauge symmetry. 
Typical examples include the tachyon-free type II $\N_4=(4,0)$ models 
at finite temperature and in the presence of non-trivial gravito-magnetic fluxes, 
which are described in great detail in the literature \cite{akpt,massivesusy,FKT,FKPT}.  
The fundamental properties of these models, 
which can lead to the resolution of the Hagedorn and the initial singularity, 
are well understood from the recent study of 
the two dimensional Hybrid models \cite{FKPT}, and are exhibited below: 
\begin{itemize}
\item
The canonical thermal ensemble is modified by turning 
on non-trivial gravito-magnetic fluxes, which lift the usual Hagedorn instabilities. 
The fluxes inject non-trivial winding and momentum charges into the thermal vacuum and 
render the mass of the would-be tachyonic states semi-positive definite. 
The tachyon-free models are equivalent to freely acting asymmetric orbifolds obtained 
by modding out with $(-1)^{F_L}\delta_0$, $F_L$ being the left-moving space-time fermion number 
and $\delta_0$ an order-2 shift along the Euclidean time cycle. Essentially, the fluxes regulate 
the contribution to the free energy of the massive string states.    
\item
Not only is the resulting spectrum of thermal masses semi-positive definite, 
but also the partition function is duality invariant under $\beta \to \beta_c^2/\beta$, 
and it is finite for all values of $\beta$. At the critical point, 
new massless thermal states appear extending the $U(1)_L$ gauge symmetry associated 
to the Euclidean time cycle to a non-Abelian $[SU(2)_L]_{k=2}$ symmetry. 
This is a purely stringy phenomenon, absent in any conventional field theory model. 
The self-dual point $\beta=\beta_c$ is realized at the so called fermionic point. 
This universal property of all such superstring models follows from 
the conformal transformation properties of the left-moving $NS$ vacuum ($h_L=-{1\over 2}$). 
\item
The extra massless states at the critical (fermionic) point have non-trivial 
momentum and winding charges so that $p_L = \pm 1$ and $p_R=0$. 
These two extra states together with the thermal radius modulus 
give rise to the $SU(2)$ enhanced symmetry. At the critical point, 
the massless states give rise to non-trivial backgrounds 
which admit a localized brane interpretation in the Euclidean.   
\item
For ${\beta/\beta_c }\gg 1$, the asymptotic behavior of the thermal partition function 
is dominated by the light thermal momentum states giving rise to
the characteristic behavior of massless thermal radiation in $d$ dimensions, 
modulo exponentially suppressed contributions from the massive string oscillator states:
\be
{Z \over V_{d-1}} ={n^* \Sigma_d\over \beta_c^{d-1}}\left({\beta_c \over \beta}\right)^{d-1}
+\;\O\left ( e ^{-\beta/\beta_c}  \right),
\ee 
where $n^*$ counts the number of effectively massless degrees of freedom; 
$\Sigma_d$ is the Stefan-Boltzmann constant for radiation and $V_{d-1}$ is the spatial volume. 
What is extremely important is that thanks to the thermal duality symmetry, the asymptotic behavior for 
${\beta/\beta_c }\ll 1$ is dual-to-thermal, as it is dominated by the light thermal winding states:
\be
{Z \over V_{d-1}} ={n^* \Sigma_d\over \beta_c^{d-1}}\left({\beta \over \beta_c}\right)^{d-1}
+\; \O\left ( e ^{-\beta_c/\beta}  \right).
\ee 
Here also, the oscillator states give exponentially suppressed contributions. 
The contribution of the massive oscillator states remains finite at the critical point, 
as the fluxes modify and effectively reduce the density of thermally excited massive oscillator states. 
In most cases, the contribution of the massive oscillator states 
never dominates over the thermally excited massless states due to asymptotic supersymmetry 
\cite{KutSeiberg,Misaligned,akpt,massivesusy,FKT,FKPT}.  
\item
This behavior indicates the appropriate, {\it duality invariant definition of the temperature} $T$, 
valid in both asymptotic thermal regimes. Defining the thermal modulus $\sigma$ 
by $ e^{\sigma}=\beta/\beta_c $, the duality invariant temperature 
is given by  $T\equiv T_c\,  e^{-\abs \sigma\abs}$. 
Thus the temperature in these configurations, 
and consequently the energy density and pressure, never exceed a critical value. 
The maximal critical temperature is given by $T_c = 1 / \beta_c$. In both asymptotic regimes
($T\ll T_c$), the partition function can be expressed in terms of the self-dual temperature as follows:
\be
{Z \over V_{d-1}} =n^* \,\Sigma_d\,T_c^{d-1}\left({T \over T_c}\right)^{d-1}
+\; \O\left ( e ^{-T_c/T}  \right).
\ee 
\end{itemize}
We conclude that the stringy thermal system has three characteristic regimes:
The two dual phases of light thermal momenta and light thermal windings, 
and a third intermediate regime corresponding to the extended symmetry point, where
vortices described by massless thermal states carrying non-trivial momentum 
and winding charges become relevant. 

The presence of the localized massless states is crucial since {\it they 
can marginally induce transitions between purely momentum and purely winding states}, 
thus driving a phase transition between the two asymptotic regimes. 
As in \cite{FKPT}, this phase transition admits a geometrical description, 
in terms of a ``T-fold'', with branes localized at the critical point gluing 
the ``momentum'' and ``winding'' spaces. This gluing mechanism 
was explicitly realized in the two dimensional Hybrid model \cite{FKT,FKPT}, 
where the partition function and its conical structure 
were determined beyond any $\alpha^{\prime}$ approximation 
at the perturbative genus-1 level. In the Hybrid model, 
the ingredients described above not only treat successfully the Hagedorn transition, 
but also they lead to non-singular thermal cosmologies in contrast to field theoretic cases. 
In this work, utilizing the fundamental ingredients and especially the branes 
sourced by the extra massless thermal states, 
we show that non-singular string cosmologies also exist in higher dimensions.

The plan of the paper is as follows.
In section \ref{transi}, we construct type II thermal vacua in arbitrary dimension $d\ge 2$, which are free of Hagedorn instabilities due to the presence of certain gravito-magnetic fluxes. We show that up to exponentially suppressed terms, the corresponding partition functions are well-approximated by the contributions of the thermally excited massless states up to the critical point. In section \ref{eff action}, we present an effective action valid in all three regimes associated to the ``thermal momentum phase'' ($\beta>\beta_c$), the ``thermal winding phase'' ($\beta<\beta_c$) and the non-geometrical $[SU(2)_L]_{k=2}$ point ($\beta=\beta_c$). At the critical temperature, additional massless thermal states source negative pressure contributions to the effective action. Cosmological solutions free of initial singularities in arbitrary dimension are exhibited in section \ref{sol}. They are compatible with perturbation theory throughout the evolution and describe bouncing cosmologies, where a phase transition connects a contracting thermal winding space-time to an expanding thermal momentum space-time. Solutions which are radiation or curvature dominated at both the very early and very late cosmological times are presented.  Finally, our results and perspectives are summarized in section \ref{conclu}. 

%%%%%%%%%%%%%%%%%%%%%%%%%%%%%%%%%%%%
%%%%%%%%%%%%%%%%%%%%%%%%%%%%%%%%%%%%

\section{Thermal duality and the Hagedorn transition}
\label{transi}

In this section we construct tachyon-free thermal configurations, 
starting with type II $\N_4=(4,0)$ models in various dimensions. 
The left-moving worldsheet degrees of freedom 
give rise to $16$ real supercharges, while the remaining right-moving supersymmetries 
are broken spontaneously via asymmetric geometrical fluxes \cite{akpt,massivesusy,FKT,FKPT}.
Geometrical fluxes generalize the Scherk-Schwarz mechanism \cite{SS} to string theory 
\cite{SSstring,RK,GeoFluxes}. 
At certain points in 
moduli space, where the moduli participating in the
breaking of the right-moving supersymmetries attain values close to the
string scale, the local gauge symmetry is enhanced to a non-Abelian one \cite{akpt,massivesusy,FKT,FKPT}. 
Finite temperature and quantum effects will stabilize these moduli at such extended symmetry points 
\cite{min,stabmod,CriticalCosmo}. 
Interesting examples include the 
two-dimensional Hybrid models, 
where the right-moving sector is characterized by unbroken 
massive spectrum (boson/fermion) degeneracy symmetry ($MSDS$) \cite{massivesusy,FKT,FKPT}.

Finite temperature is introduced along with non-trivial gravito-magnetic fluxes, 
threading the Euclidean time cycle together with other cycles 
responsible for the breaking of the right-moving supersymmetries \cite{akpt,massivesusy,FKT,FKPT}. 
These fluxes inject into the thermal 
vacuum non-trivial momentum and winding charges and lift the Hagedorn instabilities of the thermal ensemble. 
To see how this occurs, recall that for special values of the fluxes, 
the model is equivalent to a freely acting asymmetric orbifold of the form $(-1)^{F_L}\delta_0$, 
where $F_L$ is the left-moving space-time fermion number 
and $\delta_0$ is a $Z_2$-shift along the Euclidean time circle \cite{akpt}. 
The genus-$1$ partition function is given by
\be
\begin{array}{ll}
&\dis \!\!\!\!\!\!\!\!\!\!\!\!\!Z  = {V_{d-1} \over (2\pi)^{d-1}}\int_{\cal F} \frac{d^2 \tau}{4\tau_2^{(d+1)/2}}  
{1\over (\eta \bar \eta)^8}~{1 \over 2}\sum_{\bar a,\bar b}~(-1)^{\bar a + \bar b}
~{{\bar\theta}[^{\bar a}_{\bar b}]^4 \over \bar \eta^4}
~\Gamma_{(10-d,10-d)}[^{\bar a}_{\bar b}]\tesp\\
&\dis \!\!\times\!\!\sum_{m_0,n_0}\left(V_8 \Gamma_{m_0,2n_0}(R_0) + O_8 \Gamma_{m_0 +\frac{1}{2},2n_0 +1}(R_0) - S_8  
\Gamma_{m_0 +\frac{1}{2} , 2 n_0}(R_0) -C_8 \Gamma_{m_0, 2n_0 +1}(R_0) \right)\!, 
\end{array}
\label{Z1}
\ee
where $V_{d-1}$ stands for the volume of the large spatial directions. 
Here $\Gamma_{(10-d,10-d)}[^{\bar a}_{\bar b}]$ denotes the asymmetrically 
half-shifted lattice associated with the compact directions, leading to the breaking of the right-moving supersymmetries at zero temperature. The last line in the integrand gives the combined effect of finite temperature and the gravito-magnetic fluxes, amounting to a thermal action on the left-movers of the worldsheet. As a result, the orbifold deviates from the conventional thermal one in sectors with odd right-moving fermion number $F_R$. At the preferred right-moving extended symmetry points, these sectors are massive with masses of order the string scale.  Sectors with even $F_R$ and in particular the initially massless bosons and fermions are thermally excited as in the canonical ensemble. Thus the deformed model does not differ appreciably from
the conventional thermal one at temperatures lower than Hagedorn. The presence of the gravito-magnetic fluxes however allows for
the existence of stable phases above the Hagedorn temperature for which the canonical thermal ensemble fails to converge.

The left-moving $O_8$ sector carries non-trivial momentum and winding charges so that 
\be
{1 \over 2}~{\rm min}\left\abs p_L^2 -p_R^2\right\abs = {\rm min}\left\abs (m_0+{1\over 2})(2n_0 +1)\right\abs={1\over 2}. 
\ee
This is just enough to produce a state that is at least holomorphically or anti-holomorphically massless, ensuring the absence of tachyons in the spectrum of thermal masses for all values of the thermal modulus $R_0$. In addition, the model remains tachyon free under all marginal deformations of the transverse dynamical moduli associated with the compact manifold. 
The partition function is finite and invariant under thermal duality symmetry, 
$R_0 \to 1/(2 R_0)$ (together with the $S_8 \leftrightarrow C_8$ interchange), with the self-dual critical point occurring at the fermionic point $R_c = 1/\sqrt{2}$. At $R_0=R_c$, additional thermal states become massless, enhancing the $U(1)_L$ gauge symmetry associated with the compact 
Euclidean time circle to a non-Abelian $[SU(2)_L]_{k=2}$ symmetry. Their masses are equal to 
\be
m^2=\left({1\over 2R_0} -R_0\right)^2. 
\ee 
At the fermionic point, the corresponding left- and right-moving momenta and the associated vertex operators are given by
\be
\label{O+-}
p_L=\pm 1,~~~p_R=0,~~~O_\pm=\psi_L^0\, e^{\pm i  X_L^0}\, O_R,
\ee 
where $O_R$ are weight $(0,1)$ right-moving operators. Such states can marginally induce transitions between purely thermal winding and purely thermal momentum states, and in addition exchange the spinor chirality, $S_8\leftrightarrow C_8$ \cite{FKPT}. As we will see, these states induce {\it a universal, non-analytic conical structure in the partition function $Z$} as a function of the thermal modulus $R_0$, irrespectively of the dimensionality of the model. 

Thermal duality implies a maximal critical temperature and the existence of two dual asymptotically cold regimes dominated by the light thermal momenta and the light thermal windings respectively. 
The regime of light momenta, $R_0 \gg R_c$, gives rise to a thermal phase with 
temperature $T=1/\beta$, where $\beta =2\pi R_0$ is the period of the Euclidean time circle. In the regime of light windings, $R_0 \ll R_c$, the vortices can be interpreted by T-duality as ordinary thermal excitations associated with a large circle of period $\tilde \beta = \beta_c^2/\beta$. The corresponding temperature is given by 
$T=1/\tilde \beta = \beta/\beta_c^2$. Thus the system at small radii is again effectively cold. The two phases are distinguished by the light thermally excited spinors: At large radii these transform under the $S_8$-Spinor of $SO(8)$, while at small radii they transform under the conjugate $C_8$-Spinor. At $R_0 = R_c$, we find the intermediate regime where extra massless thermal states appear and enhance the local gauge symmetry
to a non-Abelian gauge symmetry. As in Ref. \cite{FKPT}, the extra massless thermal states give rise to genus-0 backgrounds, present when $R_0=R_c$, admitting a Euclidean brane interpretation \cite{FKPT}. Since transitions between purely momentum and purely winding states can occur in the presence of 
such condensates, the branes ``glue together'' the light momentum and light winding regimes.

Defining the thermal modulus $\sigma$ by $R_0/R_c = e^\sigma$, the duality invariant expression for the temperature, valid in both 
the winding and momentum phases, is given by
\be
T=T_c~ e^{-\abs \sigma \abs}, ~~~T_c={1 \over \beta_c}= {1 \over \sqrt{2}\pi},
\ee
attaining a maximal critical value $T_c$ at the self-dual point $\sigma=0$. As a result, the energy density and pressure in these models are bounded, never exceeding certain maximal values. 

To identify further universal features concerning the thermal effective potential, it is illuminating to compare two models of different dimensionality, where the mechanism leading to the resolution of the Hagedorn singularity is transparent:

\noindent 
$\bullet$ A $d=2$ Hybrid model, where 
the first line of Eq. (\ref{Z1}) is given by
\be
{1\over (\eta \bar \eta)^{8}}~{1 \over 2}\sum_{\bar a,\bar b}~(-1)^{\bar a + \bar b}~{{\bar\theta}[^{\bar a}_{\bar b}]^4 \over \bar \eta^4}
~\Gamma_{(8,8)}[^{\bar a}_{\bar b}]~= {\Gamma_{E_8}(\tau) \over \eta^8}\left(\bar V_{24}- \bar S_{24}\right),
\ee
exhibiting holomorphic/anti-holomorphic factorization and right-moving $MSDS$ structure, as exemplified by the identity $\bar V_{24}-\bar S_{24}=24$. This model was analyzed extensively in \cite{FKT,FKPT}.

\noindent 
$\bullet$ A $d$-dimensional model, where the breaking of the right-moving supersymmetries occurs via the coupling to the right-moving space-time fermion number $F_R$  of the momentum and winding charges associated to a single factorized cycle, whose radius we denote by $R_9$ \cite{akpt}:
\be
\begin{array}{ll}
\dis \!\!\!{1\over (\eta \bar \eta)^{8}}&\dis \!\!\!{1 \over 2}\sum_{\bar a,\bar b}~(-1)^{\bar a + \bar b}
~{{\bar\theta}[^{\bar a}_{\bar b}]^4 \over \bar \eta^4}
~\Gamma_{(10-d,10-d)}[^{\bar a}_{\bar b}]= {\Gamma_{(9-d,9-d)} \over (\eta \bar \eta)^{8}}\desp\\
&\dis \times\sum_{m_9,n_9}\left(\bar V_8 \Gamma_{m_9,2n_9}(R_9) + \bar O_8 \Gamma_{m_9 +\frac{1}{2},2n_9 +1}(R_9) - \bar S_8  
\Gamma_{m_9 +\frac{1}{2} , 2 n_9}(R_9) -\bar C_8 \Gamma_{m_9, 2n_9 +1}(R_9) \right) .
\end{array}
\ee
The breaking of the right-moving supersymmetries has a similar algebraic structure to the
temperature breaking that acts on the left-moving characters, see Eq. (\ref{Z1}). The Euclidean partition function is invariant under the $R_9 \leftrightarrow R_0$ exchange\cite{akpt}. The fermionic point $R_9=R_c$ corresponds to an extended symmetry point,  with the $U(1)_R$ gauge symmetry associated with the $X^9$-cycle getting enhanced to $[SU(2)_R]_{k=2}$.  As we already stated, finite temperature and quantum effects give rise to an effective potential that stabilizes dynamically the value of the $R_9$ modulus at the extended symmetry (fermionic) point \cite{min,stabmod}. 
This behavior is {\it drastically different than that of the thermal $R_0$ modulus}, where the back-reaction in the Lorentzian drives a cosmological evolution towards smaller temperatures. 
All other spectator moduli are either stabilized at extended symmetry points or frozen at values of order unity \cite{stabmod,CriticalCosmo}. The string coupling is taken to be sufficiently weak.  As we will show, it remains smaller than a critical value during the induced cosmological evolution.  
Despite the lack of $MSDS$ structure in the higher dimensional cases,  the massive states are characterized by asymptotic supersymmetry, thanks to the asymmetric nature of  the left- and right-moving supersymmetry breakings. This fact also explains the absence of tachyons from the spectrum of thermal masses \cite{KutSeiberg,akpt}.

In the two dimensional Hybrid model, the genus-$1$ partition function takes a very simple form, which makes transparent  the conical structure of the thermal partition function. The expression can be derived 
beyond any $\alpha^{\prime}$ approximation, thanks  to the unbroken \emph{MSDS} symmetry characterizing the right-moving sector, and it is given by \cite{FKT,FKPT}:
\be \label{Zh}
{Z_{\rm Hybrid} \over V_1}=24\times \left(R_0 + {1 \over 2R_0}\right)-24 \times \left| R_0 - {1 \over 2R_0}\right|
=24\sqrt{2}~e^{-|\sigma|}.
\ee
The essential feature is a discontinuity in the first derivative of $Z_{\rm Hybrid}$ as a function of the
thermal modulus $\sigma$, signaling a phase transition between the two dual ``momentum'' and ``winding'' regimes. The discontinuity occurs at the self-dual, fermionic point $\sigma=0$, and it is sourced by the $24$ complex lowest mass states in the $O_8\bar V_{24}$-sector, which become massless precisely at this point. 
 Due to the unbroken $MSDS$ right-moving structure, there are exact cancellations between
the massive fermionic and massive bosonic oscillator states, and so both regimes at $\abs \sigma \abs >0$ comprise phases where the equation of state is effectively that of massless thermal radiation in two dimensions. Despite the cancellations in the massive sector, stringy behavior survives, with states carrying both momentum and winding charges becoming massless,  inducing a phase transition at the critical point $\sigma =0$. 

In terms of the duality invariant temperature, the partition function is simply given by 
\be
\label{ZHy}
{Z_{\rm Hybrid} \over V_1}= {48 \pi}\,T,
\ee
attaining a maximal value at the critical temperature $T_c$. The existence of this maximal  temperature
is the crucial difference from field theory thermal models where the temperature is unbounded.  
The numerical coefficient in Eq. (\ref{ZHy}) is easily understood as follows. In $d$ dimensions, the partition function corresponding 
to massless thermal radiation can be written as 
\be
{Z \over V_{d-1}}= {n^* {\Sigma}_d}~T^{d-1},
\ee 
where $n^*$ is given in terms of the numbers of the initially massless bosons and fermions and $\Sigma_d$ is the Stefan-Boltzmann constant: 
\be
n^* = n^B + n^F\,  {2^{d-1}-1 \over 2^{d-1}}, ~~~~~~ \Sigma_d={\Gamma(d/2)\over \pi^{d/2}}\, \zeta (d).
\ee
Taking into account that initially there are $8\times 24$ massless bosons and 
$8\times 24$ massless fermions in the two dimensional Hybrid model and that $\Sigma_2={\pi / 6}$, we obtain $n^*_H \Sigma_2 = 48\pi$. 

We now proceed to analyze the $d$-dimensional model. The non-analytic, conical structure of the partition 
function occurs at the fermionic point $R_0 =R_c$. We can identify it by computing $Z$ for $R_0 > R_c$ and also for $R_0 < R_c $, and then utilize thermal duality to connect the two regimes so
as to obtain an expression valid for all radii (see Refs\cite{FKT,FKPT} for detailed discussions concerning the Hybrid model and also \cite{DLS} for the two-dimensional Heterotic strings).  
For $R_0 > R_c$, we Poisson resum over the momentum quantum number $m_0$,
and map the integral over the fundamental domain to an integral over the strip \cite{FtoStrip} involving
the $(\tilde m_0, n_0)=(2\tilde k+1,0)$ orbits only. In the strip representation, the winding contributions and 
in particular those of the $O_8$ and $C_8$ sectors to the fundamental domain integral, 
are mapped to $V_8$- and $S_8$-sector momentum contributions in the UV region of the strip,
$\tau_2 < 1$. Mapping the integral over the fundamental domain to an integral over the strip as in \cite{FKT,FKPT}
(see also \cite{FtoStrip}), we obtain
\be
\begin{array}{ll}
\dis {Z \over V_{d-1}}=&\!\!\!\dis 2R_0\sum_{\tilde k=0}^{\infty}\int_{||} \frac{d^2 \tau}{4(2\pi)^{d-1}\tau_2^{1+d/2}}~
e^{-(2\tilde k+1)^2{\pi R_0^2\over \tau_2}}~ {\theta_2^{4}\over \eta^{12}}\tesp\\
&\dis \!\!\!
\times{\Gamma_{(9-d,9-d)}\over \bar \eta^8}\sum_{m_9,n_9}\left(\bar V_8 \Gamma_{m_9,2n_9} + \bar O_8 \Gamma_{m_9 +\frac{1}{2},2n_9 +1} - \bar S_8 
\Gamma_{m_9 +\frac{1}{2} , 2 n_9} -\bar C_8 \Gamma_{m_9, 2n_9 +1} \right)\!\bigg\abs_{R_9=R_c},
\end{array}
\label{Zstrip}
\ee
where the moduli associated with the $\Gamma_{(9-d,9-d)}$ lattice are taken to be
of order unity. 

For large $R_0$, we split the contributions to the integral in two pieces: 
($i$) the contribution of the thermally excited massless bosons and fermions 
and ($ii$) the contributions of the massive states. 

($i$){\it Massless contributions:} 
$~$\\
The contribution of the initially massless bosons and fermions is given by
\be
\begin{array}{ll}
\dis I_{\rm massless}&\!\!\!\dis =2R_0\sum_{\tilde k=0}^{\infty}\int_{0}^{\infty} \frac{d \tau_2}{2(2\pi)^{d-1}\tau_2^{1+d/2}}~
e^{-(2\tilde k+1)^2{\pi R_0^2\over \tau_2}}~16\times (8+2),\desp\\
&\dis \!\!\!=8\times (8+2)~{2^d-1\over 2^{d-1}}~{\Gamma(d/2)~\zeta(d) \over \pi^{d/2}~ \beta^{d-1}} 
= n^* \Sigma_d{1 \over \beta^{d-1}}~.
\end{array}
\ee

($ii$){\it Massive contributions:}
$~$\\
To determine the contribution of the massive states, we must first compute the integral over $\tau_1$, which imposes level matching. The integral over $\tau_2$ gives the final result in terms of Bessel functions. 
Due to the presence of the gravito-magnetic fluxes, there are alternating signs between the contributions of states in the $F_R$-even and $F_R$-odd sectors (right-moving bosons and right-moving fermions).
The massive contributions become 
\be
I_{\rm massive}={2 \over (2\pi)^{d/2}\beta^{d/2 -1}}~ \sum_i \sum_{\tilde k = 0}^{\infty}~(-1)^{F_R}~{\abs m_i \abs^{d/2} \over (2\tilde k+1)^{d/2}}~K_{d/2}\left((2\tilde k+1)\beta m_i\right),
\ee
where the first sum is over individual degenerate boson/fermion pairs, having the same right-moving fermion number. Thanks to the $(-1)^{F_R}$ alternating signs, the effective density of states gets reduced drastically, as compared to the canonical thermal ensemble. This is a signal of right-moving asymptotic supersymmetry, replacing the exact right-moving $MSDS$ structure of the Hybrid models. The massless sector on the other hand contributes identically as in the canonical thermal ensemble. For $\beta \gg \beta_c$, the arguments of the Bessel functions are large, leading to an exponential suppression of the contributions of all massive states. In this regime, we recover the characteristic behavior of massless thermal radiation in $d$ dimensions:
\be
{Z \over V_{d-1}} \sim I_{\rm massless}={n^* \Sigma_d\over \beta_c^{d-1}}\left({\beta_c \over \beta}\right)^{d-1}.
\ee 
By thermal duality, the behavior for $\beta \ll \beta_c$ is dual-to-thermal, yielding    
\be
{Z \over V_{d-1}} \sim {n^* \Sigma_d\over \beta_c^{d-1}}\left({\beta \over \beta_c}\right)^{d-1}.
\ee 
This T-dual result for $R_0 \ll R_c$ can also be obtained if we first Poisson resum over the winding number $n_0$ and utilize the momentum-unfolding to map the integral over the fundamental domain to an integral over the strip. 

To complete the discussion, we have to analyze the behavior of the partition function in the intermediate regime when $R_0$ is close but still larger than the fermionic point $R_c$.
In this case, we must examine the contribution of the $\tau_2 \to 0$ region to the integral (\ref{Zstrip}), due to the exponential growth in the density of massive states in each sector of definite $F_R$ parity separately. The individual contributions of the massive states are larger in this region. 
However, as we will see the contribution from this region is drastically reduced due to right-moving asymptotic supersymmetry, whose origin is the insertion of the $(-1)^{F_R}$ phase. 
This property also explains the absence of physical tachyons from the spectrum of thermal masses.  
 The asymptotic $\tau_2 \to \infty$ region is dominated by the lightest string states giving rise to the thermal massless radiation contribution.  As we show below this contribution will turn out to be the dominant one.

To proceed further we need to determine the $\tau \to 0$ limit of the integrand in Eq. (\ref{Zstrip}). To this end, 
it is convenient to rewrite the integrand in terms of shifted lattices $\Gamma[^h_g]$:
\be
\label{Zstriplat}
\begin{array}{ll}
\dis {Z \over V_{d-1}}=&\!\!\!\dis -2R_0\sum_{\tilde k=0}^{\infty}\int_{||} \frac{d^2 \tau}{4(2\pi)^{d-1}\tau_2^{1+d/2}}~ e^{-(2\tilde k+1)^2{\pi R_0^2\over \tau_2}}~ {\theta_2^{4}\over \eta^{12}}\desp\\
&\dis \!\!\!\times~{\Gamma_{(9-d,9-d)}\over {\bar \eta }^8}~{1\over \bar \eta^{4}}\left.\left(\Gamma[^1_1]~\bar \theta_3^{4}-\Gamma[^1_0]~\bar \theta_4^{4}-\Gamma[^0_1]~\bar\theta_2^{4}\right)\right\abs_{R_9=R_c},
\end{array}\ee
where 
\be
\Gamma[^h_{\tilde g}](R)={R \over \sqrt{\tau_2}}\sum_{\tilde m, n}
e^{-\frac{\pi R^2}{\tau_2}|(2\tilde m+\tilde g) +(2n+h)\tau|^2}.
\ee
We then apply the modular transformation $\tau \to \tilde \tau =-1/\tau$ to the following expression
\be
\begin{array}{c}
\!\!\!\!\!\!\!\!\!\!\!\!\!\!\!\!\!\!\!\!\!\!\!\!\!\!\!\!\!\!\!\!\!\!\!\!\!\!\!\!\!\!\!\!\!\!\!\!\!\!\!\!\!\!\!\!\!\!\!\!\!\!\!\!\!\!\!\!\!\!\!\!\!\!\!\!\!\!\!\!\!\dis {1 \over \tau_2^{d/2-1}}{\theta_2^{4}\over \eta^{4}}~
{\Gamma_{(9-d,9-d)}\over (\eta \bar \eta)^8}
~{1\over \bar \eta^{4}}\left(\Gamma[^1_1]~\bar \theta_3^{4}-\Gamma[^1_0]~\bar \theta_4^{4}-\Gamma[^0_1]~\bar\theta_2^{4}\right)=\desp\\
\;\;\;\;\;\;\;\;\;\;\;\;\;\;\;\;\;\;\;\;\;\;\;\;\;\;\;\;\;\;\;\;\;\;\;\;\;\dis {1 \over \tilde\tau_2^{d/2-1}}\left[{\theta_4^{4}\over \eta^{4}}~
{\Gamma_{(9-d,9-d)}\over (\eta \bar \eta)^8}
~{1\over \bar \eta^{4}}\left(\Gamma[^1_1]~\bar \theta_3^{4}-\Gamma[^1_0]~\bar \theta_4^{4}-\Gamma[^0_1]~\bar\theta_2^{4}\right)\right]\!(\tilde \tau),
\end{array}
\ee
which appears in the integrand of Eq. (\ref{Zstriplat}). The last expression can be expanded in powers of 
$\tilde q =e^{2\pi \tilde \tau}$, in the limit $\tilde q\to 0$. For $R_9=R_c$, we obtain the following leading terms 
\be
-{1 \over \tilde \tau_2^{d/2-1}}\left(\tilde q^{-{1\over 2}} ~
-~8~\right)\left(8~+~2~+~2~\tilde q^{{1\over 2}}~{\bar {\tilde q}}^{-{1\over 2}}\right).
\ee
Essentially, only the left/right level matched terms contribute due to the integration over $\tau_1$. Keeping these only, we get
\be
{1 \over \tilde \tau_2^{d/2-1}}~8\times\left(8~+2\right)=
 \tau_2^{d/2-1}8\times\left(8~+2\right).
\ee 
The absence of exponential growth in this factor is the signal of asymptotic supersymmetry, as was already stated before. 
Notice that for $d>2$, this factor goes to zero as a power law. So, the contribution from this region is estimated to be
\be
-8\times\left(8~+2\right)~{R_c \over(2\pi)^{d-1}} ~\int_{t^*}^{\infty} \frac{d \tilde \tau_2}{\tilde \tau_2^{1+d/2}}~
e^{-{\pi R_c^2\tilde \tau_2}}~~\sim ~{\rm \cal O} \left(~{e^{-\pi L}\over L^{d+2\over 2}}~\right),
\ee
where $t^*={L\over R^2_c} $ is a sufficiently large cutoff, $L \gg 1$.
The overall contribution is thus exponentially suppressed.

Therefore the behavior of $Z$ for $R_0 > R_c$ is controlled by the thermally excited initially massless states everywhere and up to the critical point. The same conclusion can be reached in the regime $R_0< R_c$ by thermal duality. Gluing the two regimes in a duality invariant way gives
\be
\label{local beha}
{Z \over V_{d-1}} = {n^*\Sigma_d \over \beta_c^{d-1}}\,e^{-(d-1)\abs \sigma \abs}=n^*\Sigma_d\,T^{d-1},
\ee
modulo the exponentially suppressed contributions in the three effective field theory regimes. This result also implies that in each of the two
thermal phases the various thermodynamical quantities enjoy the standard monotonicity properties as functions of the temperature, with the
specific heat being positive up to the critical point.  

To conclude, the above result reveals a universal conical structure at $\sigma =0$, irrespectively of the dimensionality of the model. All of the above manipulations, including thermal duality, amount to approximating the last line of equation ($\ref{Zstrip}$) with a factor of order unity:
\be
{Z \over V_{d-1}}=2R_ce^{\abs \sigma\abs}~\sum_{\tilde k=0}^{\infty}\int_{||} \frac{d^2 \tau}{4(2\pi)^{d-1}\tau_2^{1+d/2}}~
{\rm exp}\left({{-(2\tilde k+1)^2\,\pi R_c^2\over \tau_2}~e^{2\abs \sigma \abs}}\right)~ {\theta_2^{4}\over \eta^{12}}~(8+2).
\ee
Thanks to the analytic properties of the left-moving characters, only the massless level
contributes in the above integral. This is the generalization of the Hybrid model result \cite{FKT,FKPT} 
to arbitrary dimensions via right-moving asymptotic supersymmetry.

%%%%%%%%%%%%%%%%%%%%%%%%%%%%%%%%%%%%%%%%%%%
%%%%%%%%%%%%%%%%%%%%%%%%%%%%%%%%%%%%%%%%%%%

\section{Effective action(s) up to genus-1}
\label{eff action}

From the thermal effective field theory point of view, there are at least three different effective actions associated to three possible  $\alpha^{\prime}$-like expansions, each being valid in one of the three characteristic regimes. Namely: 
\be
{R_0\over R_c} \gg 1~~  (\sigma \gg 0),\qquad {R_c\over R_0} \gg 1~~(-\sigma \gg 0),\qquad\left \abs {R_0\over R_c} -{R_c  \over R_0}\right\abs\ll 1~~ (\sigma \simeq 0).
\ee
Taking into account the behavior of the thermal partition function in each regime, the thermal duality symmetry, as well as the branes which glue the dual ``momentum'' and  ``winding'' phases, we will construct a $d$-dimensional effective cosmological action valid in all regimes simultaneously and derive the associated equations of motion. In the Lorentzian, the branes are spacelike, appearing at any time when the temperature reaches its critical maximal value $T_c$. As we will argue, these source localized negative pressure contributions to the effective action. 

%%%%%%%%%%%%%%%%%%%%%%%%%%%%%%%%%%%%%%%%%%%

In the two asymptotic regimes ($\sigma \rightarrow \pm\infty$) dominated by the light thermal momenta and the light thermal windings respectively, the effective action admits the well known sigma-model descriptions, defined via the corresponding $\alpha^{\prime}$-expansions. Thanks to  thermal duality, both asymptotic regimes can be simultaneously described by {\it a unique expansion in terms of the duality invariant temperature}  $T= T_c\, e^{-\abs \sigma \abs}$.  All thermodynamical quantities such as the temperature, the energy density and the pressure are given in terms of manifestly duality invariant expressions involving the absolute value of the thermal modulus $\abs \sigma \abs$.  In the Euclidean, the regime  $\abs \sigma \abs \rightarrow 0$ is well described in terms of the $[SU(2)_L]_{k=2}$ CFT associated with the fermionic extended symmetry point. In this regime, we have to include the contributions of the extra massless thermal states, responsible for the phase transition,  both at the genus-0 and genus-1 levels. The genus-0 contributions admit a brane interpretation with a tension determined by the allowable non-trivial backgrounds of the extra massless thermal scalars, $\partial_{\hat\mu}\varphi^I\ne 0$, where the gradients are  along the directions transverse to Euclidean time. 

During the cosmological evolution, the thermal modulus $\sigma$ acquires non-trivial time-dependence, $\sigma(\tau)$. Since all fields are functions of $\abs \sigma(\tau) \abs$, their second time-derivatives may give rise to localized singular terms proportional to 
\be
\label{sigmatau}
\delta \left(\sigma(\tau)\right)= \sum_i\, {d\tau\over d\sigma}~\delta (\tau-\tau_i),
\ee
where the temperature reaches its critical value at times $\tau_i$ $(i=1,\dots,n)$ so that $\sigma(\tau_i)=0$. As we will see, these singularities are naturally resolved by the presence of the spacelike branes, localized at the times $\tau_i$.  

The relevant representations of the winding-like field theory are space-time left-moving Vectors $V_8$ and space-time anti-Spinors $C_8$. On the other hand, in the momentum-like field theory, the relevant operators are the left-moving Vectors $V_8$ and space-time Spinors $S_8$. At the branes, the theory is self-dual and both the Spinor and anti-Spinor representations coexist together with extra massless states with non-trivial momentum and winding charges, $(p_L,p_R)=(\pm 1,0)$, triggering the transition of the winding-like field theory based on $V_8-C_8$ to the momentum-like one based on $V_8-S_8$.

The above ingredients lead to an effective $d$-dimensional dilaton-gravity action (up to the genus-1 level), 
able to describe simultaneously and in a consistent way the
three regimes:
\begin{align}
\label{action}
	\S = \S_0 + \S_1 + \S_{\textrm{brane}},
\end{align} 
where
\begin{align}
	&\S_0 = \int{d^d x\, e^{-2\phi}\,\sqrt{-g}\left(\frac{1}{2}\,{\cal R}+2(\nabla\phi)^2\right)},\nonumber\\
	&\S_1 = \int{d^d x\, \sqrt{-g}\, P},\label{eq33}\\
	&\S_{\textrm{brane}} =-  \sum_i     \int d^dx\, \sqrt{g_\bot} \,e^{-2\phi}\,\kappa_i\,\delta (\tau-\tau_i).\nonumber
\end{align}
- $\S_0$ is the genus-0 dilaton-gravity action written in the string frame. \\
- $\S_1$ is the genus-1 contribution of the thermal effective potential $-P$. \\
-  $\S_{\rm brane}$ is the spacelike brane contribution at the phase transition giving rise to localized negative pressure. It is sourced 
by the additional massless scalars $\varphi^I$ from the $O_8\bar V_8$ and $O_8\bar O_8$ sectors at the extended symmetry point $\sigma=0$. These extra 
massless states parametrize a manifold, which up to discrete identifications, takes the form of a coset space,
\be
\label{extramanifold}
{\cal M}^{2q}(\varphi^I)={SO(2, q) \over SO(2) \times SO(q)}~ ,
\ee
of real dimension $2q$ which depends on the number of massless states coming 
from the right-moving (non-supersymmetric) sector. In a suitable parametrization, the associated K\"ahler  potential is
given by 
\be
K=-\ln\Big((Y^0+\bar Y^0)^2-(Y^\alpha+\bar Y^\alpha)^2\Big)\; , ~~\alpha=1,\dots,q-1,
\ee
with the metric given in terms of the holomorphic and anti-holomorphic derivatives of $K$:
$$
ds^2({\cal M}^{2q})=\partial_{\alpha}\partial_{\bar \beta}K~dY^{\alpha} dY^{\bar \beta}=K_{\alpha  \bar \beta}~dY^{\alpha} dY^{\bar \beta}\, .
$$
The smallest possible value for $q$ is $10-d$, which  occurs when there is no extended
right-moving gauge symmetry. In the models of \cite{akpt} described in
detail in section \ref{transi}, $q=2+(10-d)$, with the gauge symmetry of the right-moving sector being extended to ${\cal H}_R=SU(2)\times U(1)^{9-d}$ of dimension $3+(9-d)=q$. 
The maximum value for $q$ is $3(10-d)$, occurring when the gauge symmetry of the right-moving sector is extended to ${\cal H}_R$ of dimension $3(10-d)$. In the simplest models with this property, ${\cal H}_R = SU(2)^{10-d}$.

\subsection{The S-brane action}

The microscopic origin of the brane term in Eq. (\ref{action}) follows from the underlying description of
the system at the extended symmetry point. 
We will be interested in obtaining homogeneous and isotropic solutions of the bulk. 
The scalars $\varphi^I$ give rise to a tree-level localized action
\be
\label{ac1}
 \S_{\textrm{brane}}= -\int d\sigma\, d^{d-1}x \, \sqrt{g_{\bot}}\, e^{-2\phi}\,g^{\hat \mu\hat \nu}\,G_{IJ}\, \partial_{\hat \mu} \varphi^I \partial_{\hat\nu} { \varphi}^J\,\delta(\sigma),
\ee
where $\hat\mu=1,\dots,d-1$, $g_\bot=\det g_{\hat\mu\hat\nu}$ and $G_{IJ}$ (or $K_{\alpha \bar\beta}$) is the metric in the field configuration space ${\cal M}^{2q}$.
The equations of motion of the scalars $\varphi^I$ take the form
\be
\label{eomphi}
2\partial_{\hat\mu}( e^{-2\phi} \sqrt{g_\bot} ~g^{\hat\mu\hat\nu} ~G_{IJ} \partial_{\hat\nu} \varphi^J )-
e^{-2\phi} \sqrt{g_\bot} ~g^{\hat\mu\hat\nu}(\partial_IG_{KJ}) \partial_{\hat\mu} \varphi^K\partial_{\hat\nu} \varphi^J \, =0\, .
\ee
 Our aim is to establish that these equations
admit non-trivial solutions which are
consistent with the homogeneity and isotropy requirements and yield the localized brane contributions in the effective action as
described by Eqs (\ref{action}) and (\ref{eq33}). To this end, it suffices that at each instant $\tau_i$ ($i=1,\dots, n$), when the temperature reaches its critical value, the induced metric  be proportional to the spatial metric:
\be
\label{hg}
h_{\hat\mu\hat\nu}\equiv G_{IJ}~\partial_{\hat\mu}\varphi^I\,\partial_{\hat\nu}\varphi^J\,={\kappa_i\over d-1} ~g_{\hat\mu\hat\nu},
\ee 
where  the $\kappa_i$'s are positive constants. When this happens, the stress tensor of the scalars is consistent with the
symmetries of the spatial metric, and therefore with homogeneity and isotropy. 
Moreover, the action (\ref{ac1}) takes the familiar form of the Nambu-Goto action for branes,
\begin{eqnarray}
 \nonumber \S_{\textrm{brane}}\!\!\!&=&\!\!\!-\sum_i \kappa_i\int d^d x\,e^{-2\phi}\, \sqrt{g_\bot}~ \delta(\tau-\tau_i)\\ 
 \!\!\!&=&\!\!\!-\sum_i (d-1)^{d-1\over 2}\kappa_i^{3-d\over 2}\int d^dx\, e^{-2\phi}\, \sqrt{\det\left(G_{IJ} \partial_{\hat\mu} \phi^I\partial_{\hat\nu} \phi^J\right)}~ \delta(\tau-\tau_i)\,,
\end{eqnarray}
where we have used Eq. (\ref{sigmatau}). Thus, $\kappa_i$ is interpreted as a brane tension.

%%%%%%%%%%%%%%%%%%%%%%%%%%%%%%%%%%%%%
To exhibit the solutions, we first
consider the following embedding of space $\Omega^{d-1}(x^{\hat\mu})$ into the field configuration space,
$
\Omega^{d-1}(x^{\hat\mu}) ~\rightarrow~{\cal M}^{2q}(\varphi^{I}):  
$
\be
\partial_{\hat\mu}\varphi^I=~ \delta_{\hat\mu}^I,~~
\hat\mu,\,I=1,\dots,d-1, ~~~{\rm and}~~~ \varphi^I={\rm const.},~~ I=d,\dots,2q.
\ee
The embedding exists provided that the dimensionality of the scalar field manifold  
is bigger or equal to the dimension of space: $2q\ge d-1$. It also implies
that the induced metric satisfies: $h_{\hat\mu\hat\nu}= G_{\hat\mu\hat\nu}$. Thus Eq. (\ref{hg}) imposes
that the spatial metric $g_{\hat\mu\hat\nu}$ and $G_{\hat\mu\hat\nu}$ are isomorphic
\be
\label{Gg}
G_{\hat\mu\hat\nu}={\kappa_i\over d-1}~ g_{\hat\mu\hat\nu}\, ,
\ee
upon the identification $\varphi^{\hat \mu}=x^{\hat \mu}$.
Under the above circumstances, the field equations of motion (\ref{eomphi}) for $I=\hat\mu \le d-1$ become:
\be
2\partial_{\hat\mu}( e^{-2\phi} \sqrt{g_\bot}) - e^{-2\phi} \sqrt{g_\bot}~ g^{\hat\sigma \hat\nu}\partial_{\hat\mu}g_{\hat\nu\hat\sigma}=0\quad \Longrightarrow\quad  \partial_{\hat\mu}\phi=0,
\ee
consistently with the homogeneity of the dilaton field $\phi$, and where we have used the identity 
 $$
 g^{\hat\sigma \hat\nu}\partial_{\hat\mu}g_{\hat\nu\hat\sigma}=2\partial_{\hat\mu}{\rm log }\sqrt{g_\bot}.
 $$
Eqs (\ref{eomphi}) for $I>d-1$ must also be satisfied.
The above discussion makes it clear that the geometrical structure of the field manifold ${\cal M}^{2q}(\varphi^I)$ is crucial,
 since it constrains the possible embeddings of $\Omega^{d-1}(x^{\hat\mu})$ into ${\cal M}^{2q}(\varphi^I)$.
We are mainly interested for the isotropic embeddings of the hyperbolic space $H^{d-1}$ and the flat space $F^{d-1}$, 
with curvature and metric given by 
\be
(i) ~H^{d-1} : k=-1\mbox{ (for $d>2$)}\, , \quad g_{\hat\mu\hat\nu}={a(\tau_i)^2\over (x^{d-1})^2}\,\delta_{\hat\mu\hat\nu}, ~~~~~
(ii) ~F^{d-1} : k=0\, , \quad g_{\hat\mu\hat\nu}=a(\tau_i)^2\, \delta_{\hat\mu\hat\nu}\, , 
\ee
where $a(\tau_i)$ is the scale factor at time $\tau_i$. 

\vspace{.3cm}
\noindent {\large $(i)$ \em Hyperbolic embedding $H^{d-1}\to {\cal M}^{2q}$}

\noindent The hyperbolic embedding turns out to be naturally realized thanks to the geometrical structure
of the field manifold ${\cal M}^{2q}$. Indeed it is sufficient to utilize the sub-manifold  
$K^q(y^I)\subset {\cal M}^{2q}(Y^I)$ with $y^I=\Re\, Y^I$ ($I=0,\dots,q-1$) non trivial and  $\omega^I=\Im \, Y^I$  fixed.
The sub-manifold  $K^q(y^I)$ naturally contains the desired $H^{q-1}$ factor:
\be
K^q\equiv H^{q-1}\times SO(1,1)={SO(1, q-1) \over SO(q-1)} \times SO(1,1)\, .
\ee
The metric on $K^q$ follows from the K\"ahler metric $K_{\alpha\bar \beta}$ and takes the form:
\be
ds^2(K^q)={-(dy^0)^2+(dy^I)^2 \over 2 r^2} + {(dr)^2 \over r^2}\, ~~~{\rm with}~~~r^2\equiv (y^0)^2-(y^I)^2.
\ee 
The imaginary parts $\omega^I$ are frozen consistently with all equations of motion.
The metric of the $(q-1)$-dimensional hyperboloid $H^{q-1}$ is obtained by writing 
\be
-(dy^0)^2+(dy^I)^2 =-dr^2+ r^2\,(dH^{q-1})^2 \, ,
\ee
which shows the explicit factorization of $K^{q}$: 
\be
\label{Kfact}
ds^2(K^{q})={1\over 2}\left(d\zeta^2+ (dH^{q-1})^2 \right)\, ,~~\zeta=\ln r\, .
\ee
Fixing further the field $\zeta$ to be constant, as allowed by the equations of motion, the embedding of the spatial hyperboloid $H^{d-1}(x^{\hat\mu})$ into $H^{q-1}(\varphi^{i})$ is automatic. This can be done, provided that $q\ge d$, $\varphi^{\hat \mu}=x^{\hat \mu}$ and
 the extra $q-d$ fields $\varphi^i, i=d,\dots q-1$ are also frozen as allowed by the equations of motion:
 \be
\label{Hmetric}
 ds^2(H^{q-1})={(d\varphi^{\hat\mu})^2\over (\varphi^{d-1})^2} +{(d\varphi^i)^2\over (\varphi^{d-1})^2}
\qquad \Longrightarrow\qquad  ds^2(H^{d-1}) ={(d\varphi^{\hat\mu})^2\over (\varphi^{d-1})^2} .
  \ee
When $q=2+(10-d)$ as in the models of \cite{akpt}, the constraint $d\le q$ implies $d\le 6$, while when $q$ takes the maximal value $3(10-d)$, we must have $d\le 7$. 
These embeddings satisfy the relation (\ref{hg}) with the tension $\kappa_i$ fixed in terms of the scale factor: $\kappa_i={(d-1)/2a(\tau_i)^2}$.
Such a tension leads to highly curved cosmological solutions with Ricci curvature of order one, as the value of $a(\tau_i)$ is fixed to be of order one by the cosmological equations (see section \ref{Thk=-1}).  
 
 To relax the constraint on the tension $\kappa_i$, we take into account discrete identifications in the field configuration space in order to obtain solutions with a non-trivial wrapping number. With suitable discrete identifications, the submanifold parametrized by the fields $\varphi^I$ ($I=1,\dots, d-1$) becomes a finite volume hyperbolic space of the form $H^{d-1}/\Gamma$, where $\Gamma$ is a subgroup of the discrete duality group $SO(2,q;\Z)$ of ${\cal M}^{2q}$. The two dimensional examples correspond to the familiar higher genus
Riemann surfaces. Homogeneous compact hyperbolic manifolds in dimension $\ge 3$  
are characterized by the property of rigidity, which
implies that there are no massless shape moduli \cite{mostow}, and are locally isotropic. The volume is determined by the radius of curvature $L$ 
and the topology of the 
manifold: ${\rm Vol}(H^{d-1}/\Gamma)=L^{d-1}e^{\alpha}$, where $\alpha$ is a constant determined by the topology and $L \sim 1$ for the metric
on the field subspace. The topological factor $e^{\alpha}$ is unbounded from above. 
Taking the spatial manifold $\Omega^{d-1}(x^{\hat\mu})$ to also be a compact hyperbolic manifold of large volume
 (and suitable topology \footnote {See e.g. \cite{starkman,krst} for discussions concerning this possibility
in the context of cosmological and other applications.}), allows for embeddings with arbitrary wrapping number $w$.
Consequently the brane action is finite, given by
\be
S^i_{\rm brane}= - {c^*\, w  \,\over a(\tau_i)^2} ~e^{-2\phi_i}~{\rm Vol}(\Omega^{d-1}) = -\kappa_i \,\int dx^{d-1} e^{-2\phi_i}\sqrt{g_\bot} \, ,
\ee 
implying
that the brane tension is
$\kappa_i = {c^*~w^2 / a(\tau_i)^2}$. 
Here $c^*$ is a factor determined by the topology. 
The wrapping number $w$ being arbitrary, the tension can be kept arbitrary.  
 
 \vspace{.3cm}
\noindent 
{\large $(ii)$ \em Flat embedding $F^{d-1}\to {\cal M}^{2q}$}
   
\noindent A way to realize the flat embedding is by utilizing a $(d-1)$-dimensional flat section  $F^{d-1}(\varphi^{\hat\mu})$ of $ {\cal M}^{2q}$. 
The isotropic embedding of flat space $F^{d-1}(x^{\hat\mu})$ into the flat section $F^{d-1}(\varphi^{\hat\mu})$ is 
defined by 
\be
\label{FlatEmb}
\partial_{\hat \nu} \varphi^{\hat\mu}=\sqrt{{2\kappa_i \over d-1}}~a(\tau_i)~\delta^{\hat\mu}_{\hat\nu}\,,
\ee 
giving rise to Eq. (\ref{hg}) and the brane action with tension $\kappa_i$.  We display below examples of such flat sections. \\
$\bullet$ For $d=2$, we can utilize the $SO(1,1)_{\zeta}$ factor of the submanifold $K^q$ parametrized by the field $\zeta$ in Eq. (\ref{Kfact}) to carry out the embedding $F^1\to {\cal M}^{2q}$. All other fields are frozen consistently with the equations of motion. \\
$\bullet$ For $d=3$, we utilize the $SO(1,1)_{\zeta}\times SO(1,1 )_{\xi}$ submanifold of $K^q$ parametrized by the fields $\zeta$ and $\xi \equiv \ln \varphi^{d-1}$, see Eq. (\ref{Hmetric}).\\
$\bullet$ For $d\ge 4$ we utilize the flat section of $H^{d-1}(u^i)$ obtained in the limit $u^{d-1}\to \infty$, together with the $SO(1,1)_{\zeta}$ factor. To this end, we define the rescaled fields $\varphi^i$ by setting $u^i\equiv u^{d-1}\varphi^i$ ($i=1,\dots, d-2$), in order to obtain:
\begin{align}
\nonumber {1\over 2}\left[d\zeta^2+ (dH^{d-1})^2\right] &={1\over 2}\left[d\zeta^2+ (d\varphi^i)^2\right]+{1\over u^{d-1}}\varphi^i\, d\varphi^idu^{d-1}+{1+(\varphi^i)^2\over 2(u^{d-1})^2}(du^{d-1})^2 \esp\\
 &=  {1\over 2}\left[d\zeta^2+ (d\varphi^i)^2\right]+{\rm \cal O}({1\over u^{d-1}})\,.
\end{align}
The field $\zeta$ and the $d-2$ fields $\varphi^i$ are utilized to realize the flat embedding of Eq. (\ref{FlatEmb}).
The $2q-(d-1)$ extra fields are frozen, consistently with the equations of motion (\ref{eomphi}), including that of $u^{d-1}$ in the limit $u^{d-1}\to \infty$. The realization of this isotropic embedding imposes the constraint $d\le q$. When $q=2+(10-d)$ as in the models of \cite{akpt}, this implies $d\le 6$, while when $q$ takes the maximal value $3(10-d)$, $d\le 7$ is required.

Finally let us note that isotropic embeddings are also possible when space is isomorphic to
the ($d-1$)-dimensional sphere $S^{d-1}$. These embeddings, the restrictions on the
dimensionality of spacetime and the structure of the bulk solutions are currently under investigation.  
This completes our discussion about the origin of the brane contributions in the effective action.

%%%%%%%%%%%%%%%%%%%%%%%%%%%%%%%%%%%%%%%%%%%

\subsection{Equations of motion}

Looking for homogeneous and isotropic cosmological solutions in dimensions $d\ge 2$, the dilaton field is a function of time only and the metric 
\be
ds^2=-N(\tau)^2d\tau^2+a(\tau)^2 (d\Omega^{d-1})^2, ~~~~~\phi =\phi(\tau)
\ee
involves the line element $(d\Omega^{d-1})^2$ of the $(d-1)$-dimensional Einstein space with curvature $k$. To derive the equations of motion, we utilize the analytic expression for the Ricci scalar curvature
\be
\label{Ricci}
{\cal R}={2(d-1)\over N^2}\left[ {\ddot a\over a}+{(d-2)\over 2} \left(H^2+{kN^2\over a^2}\right)-H {\dot N\over N}\right]\quad \where \quad H\equiv {\dot a\over a}.
\ee
The pressure $P$ is determined by the genus-1 Euclidean path integral $Z$ (the thermal partition function) as
\be
P=T_c\, e^{-\abs \sigma \abs}\,  {Z(\abs \sigma \abs ) \over V_{d-1}}\, .
\ee
It is important to note that in the sigma-model frame, $P$ is a function of the thermal modulus  $\abs \sigma \abs$ only and there is no dependence on the dilaton field $\phi$.

The above considerations  lead us to the following equations of motion:

{\bf (i)}  {\it $N$-equation}
\be
{1\over2}(d-1)(d-2) \left( H^2+ k{N^2\over a^2} \right) =2(d-1)H\dot \phi-2\dot\phi^2+e^{2\phi}N^2\rho,
\ee
where the energy density is given by\footnote{To derive $\rho$, we make use of ${\delta[N(\tau')P(\abs\sigma(\tau')\abs)]\over \delta N(\tau)}=\delta(\tau'-\tau)P+N{\partial P\over \partial\abs\sigma\abs}{\delta\abs\sigma(\tau')\abs\over \delta N(\tau)}$ where ${\delta\abs\sigma(\tau')\abs\over \delta N(\tau)}={\delta(\tau'-\tau)\over N}$, as follows from the fact that $\beta_ce^{\abs\sigma(\tau)\abs}{dx^0\over d\tau}\equiv N(\tau)$.}
\be
\rho= -P-{\partial P\over \partial\abs\sigma\abs}=-T_c\, e^{-\abs \sigma \abs}\,  {\partial\over \partial \abs \sigma \abs}\left( {Z(\abs \sigma \abs)\over V_{d-1}}\right).
\ee

{\bf (ii)}  {\it $a$-equation}
\be
\begin{array}{c}
\dis \!\!\!\!\!\!\!\!\!\!\!\!\!\!\! \!\!\!\!\!\!\!\!\!\!\!\!\!\!\! \!\!\!\!\!\!\!\!\!\!\!\!\!\!\!(d-2){\ddot a\over a}+{1\over2}(d-2)(d-3)  \left( H^2+ k{N^2\over a^2} \right)-(d-2)H{\dot N\over N}=\esp \\
\dis  \,\,\,\,\,\,\,\,\,\,\,\,\,\,\, \,\,\,\,\,\,\,\,\,\,\,\,\,\,\, \,\,\,\,\,\,\,\,\,\,\,\,\,\,\,2\ddot\phi+2(d-2)H\dot\phi-2\dot\phi^2-2\dot\phi{\dot N\over N}-e^{2\phi}N^2P + \sum_i    \kappa_i N\, \delta (\tau-\tau_i).
\end{array}
\ee

{\bf (iii)}  {\it $\phi$-equation}:
\be
\begin{array}{c}
\dis \!\!\!\!\!\!\!\!\!\!  \!\!\!\!\!\!\!\!\!\!\!\!\!\!\! \!\!\!\!\!\!\!\!\!\!\!\!\!\!\! \!\!\!\!\!\!\!\!\!\!\!\!\!\!\!  \!\!\!\!\!\!\!\!\!\!\!\!\!\!\!  \!\!\!\!\!\!\!\!\!\!\!\!\!\!\! \!\!\!\!\!\!\!\!\!\!\!\!\!\!\! \!\!\!\!\!\!\!\!\!\!\!\!\!\!\! \ddot\phi+(d-1)H\dot\phi-\dot\phi^2-\dot\phi{\dot N\over N}=\esp\\
\dis \,\,\,\,\,\,\,\,\,\,\,\,\,\,\, \,\,\,\,\,\,\,\,\,\,\,\,\,\,\, {d-1\over 2}\left({\ddot a\over a}+{d-2\over 2}  \left( H^2+ k{N^2\over a^2} \right) - H{\dot N\over N}\right)-{1\over2} \sum_i    \kappa_i  N\,\delta (\tau-\tau_i). 
\end{array}
\ee
It is useful to disentangle the second derivatives of the scale factor and dilaton field in the last two equations. Of particular interest is the linear combination which leads

{\bbf (ii)$^{\prime}$} {\it trace equation}
\be
\dot\phi^2-{d-1\over 2}\left({\ddot a\over a}-H{\dot N\over N}\right) -{1\over 4}(d-1)(d-2)\left( H^2+ k{N^2\over a^2} \right)= e^{2\phi}N^2\big[\rho-(d-1)P\big]
\ee
and shows that the second derivative of the scale factor is always finite. In particular, this implies that $\dot a$ is continuous, 
even at the branes localized at $\tau=\tau_i$. On the contrary, another equation for the dilaton is  
 
{ \bbf (iii)$^{\prime}$} {\it $\phi$-equation modulo trace equation}
\be
2\ddot \phi-4\dot\phi^2+2(d-1) H\dot\phi -2 \dot\phi{\dot N\over N}= e^{2\phi}N^2\big[(d-1)P-\rho\big]-\sum_i\kappa_i  N\,\delta (\tau-\tau_i),
\ee
which shows that the first time-derivative of the dilaton is discontinuous across $\tau_i$. This discontinuity is resolved by the presence of a spacelike brane, whose tension must satisfy
\be
\label{kphi'}
\kappa_i=2\, {\dot\phi(\tau_{i-})-\dot\phi(\tau_{i+})\over N(\tau_i)}.
\ee
 This resolution of the discontinuity via branes provides a novel mechanism in obtaining non-singular bouncing cosmologies, which in fact remain in a perturbative regime throughout the evolution. Such dynamical behavior is induced by the presence of spacelike branes providing localized negative contributions to the pressure
\be
 P_B=-\sum_i e^{-2\phi(\tau_i)}\,\kappa_i\, \delta(\tau_i)\, ,
\ee
thus evading the conditions set by the standard singularity theorems on realizing non-singular bouncing cosmologies \cite{HE}.

Before solving the above cosmological equations, it is important to stress that these give rise to an integrable relation of fundamental interest, namely the Entropy conservation equation:

{\bf (s)  \it Entropy equation}
\be 
(\dot \rho+\dot P)+ \big( (d-1)H +\abs \dot \sigma\abs \big) (\rho+P)=0.
\ee
Integrating once, we obtain a quantity $S$, which we may interpret as the conserved thermal entropy in a comoving cell of volume $a^{d-1}$:

{\bbf (s)$^{\prime}$}  {\it Entropy conservation}
\be
\label{entro}
{a^{d-1}\over T} ( \rho +P )=S.
\ee 
Since the energy density $\rho$ and the pressure $P$ are bounded, attaining their maximal values at the critical temperature $T_c$, the scale factor $a$ is bounded from below, acquiring its minimal value at the 
critical point. Thus the big-bang singularity $a =0$ of general relativity is avoided in all such Hagedorn-free string
models. The constant thermal entropy $S$ can be computed 
in the asymptotic regime where the system is radiation dominated and the dilaton is constant:
\be
\label{rad}
\dis \rho\sim (d-1) P\sim (d-1)n^* \Sigma_d\, T^d\quad \mbox{as}\quad \abs \sigma\abs\rightarrow \infty\qquad \Longrightarrow\qquad S=d\, \gamma_{\infty}^{d-1}\, n^*\Sigma_d,
\ee
where the constant $\gamma_\infty $ is given in terms of the asymptotic values of $a$ and $T$, 
$\gamma_{\infty} = \lim_{\abs\sigma\abs\to \infty} aT$.

As we demonstrated, the stringy thermal system is effectively radiation dominated in all regimes. As a result, an important consequence of the applicability of the entropy conservation in all regimes is that $\gamma=aT$ is almost constant everywhere and  up to the critical temperature. The largeness of the entropy observed at late cosmological times implies that the size of the Universe at the critical point is already large. This fact guarantees the validity of our perturbative approach as we will see later, and also, the connection of the so called entropy and oldness problems of standard Big/Bang cosmology.

%%%%%%%%%%%%%%%%%%%%%%%%%%%%%%%%%%%%%%%%
%%%%%%%%%%%%%%%%%%%%%%%%%%%%%%%%%%%%%%%%

 \section{Stringy non-singular cosmologies}
\label{sol}
 
In this section, we exhibit non-singular cosmological solutions in various dimensions. 
We will show that the mechanism for the resolutions of both the Hagedorn and the initial singularity problems are generic in all  space-time dimensions. This result follows from the universal properties of the partition function discussed in detail in sections 2 and 3. Namely:
\begin{itemize}
\item The thermal partition function has a conical singularity as a function of the thermal modulus $\sigma$ at the critical point $\sigma=0$, irrespectively of the space-time dimension. This implies a phase transition between the light-momentum and light winding-effective field theory regimes. At the phase transition, spacelike branes appear connecting the two asymptotic regimes.  
\item Across the branes, the behavior of the scale factor and dilaton field is such that  $\dot a$ is continuous, while there is a discontinuity in $\dot\phi$ related to the brane tension, see Eq. (\ref{kphi'}). 
\end{itemize}
Two-dimensional examples have already been exhibited in Ref. \cite{FKPT} in the framework of the Hybrid models, where the contributions of the massive modes to the partition function $Z$ cancel exactly thanks to the $MSDS$ structure characterizing the right-moving sector.  In the higher dimensional models presented in section \ref{transi},  the exact structure of the partition function is more involved. However, knowing the asymptotic behavior of $Z$ as $\sigma\to \pm \infty$, as well as the asymptotic density of the right-moving states, allow us to determine the essential thermodynamical properties for all temperatures  $T \le T_c$. Indeed, thanks to the asymptotic right-moving supersymmetry characterizing the models, we can well-approximate $Z$ with the contribution of the thermally excited massless states in both the $\sigma>0$ and the $\sigma<0$ regimes, as was explained in section \ref{transi}. Thus, in what follows we set
\be
\label{rad'}
\rho = (d-1) P= (d-1)\,n^* \Sigma_d\, T^d =(d-1)\,n^* \Sigma_d\, T_c^d\, e^{-d\abs \sigma \abs}   .
\ee
This last expression captures the essential conical structure of the higher dimensional cases, 
generalizing the two dimensional Hybrid results. It allows us to obtain   analytic cosmological solutions via spacelike branes, as in the two-dimensional case. The conical structure arises from matching the dual momentum and winding regimes, which admit distinct effective field theory descriptions. The intermediate $\sigma=0$ regime gives rise to the spacelike branes.  

The brane interpretation gives us the possibility to look for branches of solutions defined in time intervals $(\tau_i,\tau_{i+1})$,  with $\sigma(\tau_i)=0$. These intervals are connected to each other by spacelike branes localized at $\tau_i$. From the world-sheet point of view the transition occurs via a condensation of massless thermal states carrying non-trivial momentum and winding charges associated to the vertex operators $O_{\pm}$ defined in Eq. (\ref{O+-}).  The existence of these additional marginal operators at  $\sigma=0$ gives rise to  the transition between the thermal winding states at $\sigma=0_-$  and the thermal momentum states at $ \sigma=0_+ $ and vice-versa.  The Universe may experience a series of such phase transitions. For a cosmological evolution to be consistent, the following constraints must be fulfilled:
\begin{enumerate}

\item The dilaton and scale factor must be continuous across the branes.
\item The scale factor must be bounded from below, $a(\tau)\ge a_c$, since $T\propto 1/a$ cannot reach values larger than the maximal temperature $T_c$.
 \item Across any brane, $\dot\phi(\tau_{i-})\ge\dot\phi(\tau_{i+})$ is required, since the discontinuities in the first time derivative of the dilaton  are resolved by a positive brane tension (see Eq. (\ref{kphi'})).  
 \item  Since the first derivative of the scale factor is smooth at the branes, with the scale factor reaching its minimum value, we must have $\dot a(\tau_i)=0$. 
\item To maintain a perturbative analysis throughout the cosmological evolution, the dilaton field must be bounded from above, $\phi(\tau)\le \phi_{c}$,  with $g_c=e^{\phi_c}$ sufficiently small. 
 \end{enumerate}
 
Using the state equation (\ref{rad'}) and the entropy conservation  {\bbf (s)$^{\prime}$}, valid throughout the cosmological evolution, we obtain   
\be
\label{aT}
aT=a(\tau_i)\,T_c\equiv \gamma_\infty,\,~~~(i=1,\dots,n).
\ee
The dilaton equation {\bbf (iii)$^{\prime}$} can be easily integrated since $\rho=(d-1)P$ in each interval, yielding
\be
\label{phi'}
\dot\phi= c_i^-~ {e^{2\phi}N\over a^{d-1}}\;\;\mbox{ for }\;\tau_{i-1}<\tau<\tau_i\; , \qquad \dot\phi= c_i^+~ {e^{2\phi}N\over a^{d-1}}\;\; \mbox{ for }\;\tau_{i}<\tau<\tau_{i+1},
\ee
where $c_i^{\pm}=c_{i+1}^{\mp}$ are constants related to the tensions (\ref{kphi'}):
\be
\label{phi-jump}
\kappa_i=2\,(c_i^--c_i^+)\, {e^{2\phi(\tau_i)}\over a(\tau_i)^{d-1}}.
\ee

To count the number of independent integration constants parametrizing the solution, we consider three independent equations, namely the $N$-equation {\bbf (i)}, the dilation equation {\bbf (iii)$^{\prime}$} and the entropy conservation relation {\bbf (s)$^{\prime}$}.  These are second order  in $\phi$ and first order in $a$,  leading to three integration constants in each branch. When we glue two solutions at $\tau_i$, the continuity of $a$ and $\phi$ across the brane, together with $\dot a(\tau_{i+})=\dot a(\tau_{i-})=0$ have to be imposed. Consequently, we are left with $2\times 3-4=2$ arbitrary integration constants.  

In the following, we display solutions characterized by a  single phase transition occurring at $\sigma(\tau_c)=0,~\tau_c\equiv 0$. These solutions are obtained under the assumption that  $T_c $ is reached only once at $\tau=\tau_c=0$. For space-time dimensions $d>2$, the isotropic and homogeneous solutions may or may not have a non-trivial spatial curvature $k$. We display solutions in both cases.

%%%%%%%%%%%%%%%%%%%%%%%%%%%%%%%%%%%%%%%%

\subsection{Bouncing cosmology with vanishing curvature, $k= 0$}
\label{Thk=0}

We first consider the case of vanishing spatial curvature ($k=0$), looking for a cosmological evolution satisfying the consistency conditions 1 to 5 listed above. As explained at the end of the previous subsection,  the complete cosmological evolution is expected to depend on two arbitrary integration constants, which we may choose to be the values of the scale factor $a_c$  and the dilaton field $\phi_c$ at the brane.  We find that the complete cosmological evolution in the sigma-model frame and in the conformal gauge, 
\be
\label{confor}
\ln {N\over a_c}=\ln {a\over a_c}=\ln {T_c\over T}=\abs \sigma\abs \,,
\ee
 takes a simple form
\be
\label{sol d}
\begin{array}{l}
\dis \sigma ={\sign(\tau)\over d-2}~\left[ {\eta_+} \ln \left(1+ {\omega a_c\abs\tau\abs \over \eta_+} \right) -{\eta_-}  \ln \left(1+{\omega a_c\abs\tau\abs\over \eta_-}  \right) \right], \desp\\
\dis \phi=\phi_c+{\sqrt{d-1}\over 2}~\left[ \ln \left(1+ {\omega a_c\abs\tau\abs\over \eta_+} \right) - \ln \left(1+ {\omega a_c\abs\tau\abs\over \eta_-} \right)  \right] ,\end{array}
\ee
where  $\eta_\pm=\sqrt{d-1}\pm 1$.  The parameter $\omega$ is proportional to the brane tension $\kappa_R$, which is responsible for the gluing:
\be
\label{omegakappa}
\omega = \kappa_R\, {d-2\over 4\sqrt{d-1}} \; , \qquad \kappa_R=2\sqrt{2(d-1)}~ \sqrt{n^* \Sigma_d}~ ~T_c^{d/2}\, e^{\phi_c}\,. 
\ee
Note that $\kappa_R$ is of order $\O(e^{\phi_c}=g_c)$, which is larger than the naive $\O(e^{2\phi_c}=g^2_c)$ expectation from Eq. (\ref{phi-jump}). In the neighborhood of the brane ($\abs\kappa_R a_c \tau\abs\ll 1$), the metric is by construction regular while the dilaton field shows a conical singularity:
\be
\begin{array}{ll}
&\dis \sigma = {\sign(\tau)\over d-2} \left({1\over \eta_-}-{1\over \eta_+}\right)(\omega a_c\tau)^2+\O\big(\abs\omega a_c\tau\abs^3\big) ={\sign(\tau)\over 16(d-1)}\, (\kappa_R a_c\tau)^2+\O\big(\abs\kappa_R a_c\tau\abs^3\big) ,\desp \\
&\dis \phi =\phi_c- {\sqrt{d-1}\over 2} \left(  {1\over \eta_-}  -  {1\over \eta_+}\right)\abs\omega a_c \tau\abs +\O\big((\omega a_c\tau)^2\big)=\phi_c- { \abs\kappa_R a_c \tau\abs\over 4}+\O\big((\kappa_R a_c\tau)^2\big).
\end{array}
\ee
The maximal value of the Ricci scalar (\ref{Ricci}) occurs at the bounce and it is given in terms of the brane tension by
\be
{\cal R}_c={\kappa_R^2 \over 4}=\O(g_c^2).
\ee
Thus both higher derivative corrections and higher genus contributions in the effective action remain small throughout the evolution, and can
be consistently neglected, provided
that the critical value of the string coupling, $g_c$, is taken to be sufficiently small.

Far from the brane ($\abs\kappa_R a_c\tau\abs\gg 1$), the dilaton is asymptotically constant, the temperature drops and the scale factor tends to infinity. The whole evolution in sigma-model frame describes a bounce, where the scale factor and temperature are smooth. In the Einstein frame however, they develop conical singularities at the brane inherited from the rescaling with the string coupling. 
The Einstein frame fields are given by
\be
(N_E, a_E, 1/T_E) :=~ e^{-{2\phi \over d-2}}~(N, a, 1/T)\,.
\ee
At very early and late times ($\abs \omega a_c\tau\abs \to + \infty$), the scaling properties of the dilaton motion and thermal contributions to the energy density in the Einstein frame and in the $N_E=1$ gauge 
are respectively $1/a_E^{2(d-1)}$ and $1/a_E^d$.  The whole cosmological evolution describes a bounce between two asymptotically radiation dominated Universes.  The fact that in the Einstein frame the scale factor, the temperature and the dilaton bounce with conical singularities follows from the localized negative contribution of the brane to the pressure, $P_B=-e^{-2\phi_c}\, \kappa_R\, \delta(\tau)$. This localized pressure is not artificially added, but its origin lies in the stringy thermal duality properties of the system.  

To complete our analysis we also display the cosmological evolution in two dimensions:
\be
\label{d=2k=0}
\begin{array}{ll}
&\dis \sigma={\sign(\tau)\over 2}\left[ {\kappa_R a_c  \abs\tau\abs\over 2}-\ln \left(1+{\kappa_R a_c  \abs\tau\abs\over 2}\right) \right],\desp\\
&\dis \phi=\phi_c-{1\over 2}\ln\left(1+{\kappa_R a_c \abs\tau\abs\over 2}\right). 
\end{array}
\ee
This solution was found in the context of the Hybrid models in Ref.\cite{FKPT}, but it is also valid in the more general tachyon-free two-dimensional thermal models. Consistently, this solution is recovered when the space-time dimension $d$ is formally treated as a real parameter, taking the limit $d\to 2$ in Eqs (\ref{sol d}) and (\ref{omegakappa}).

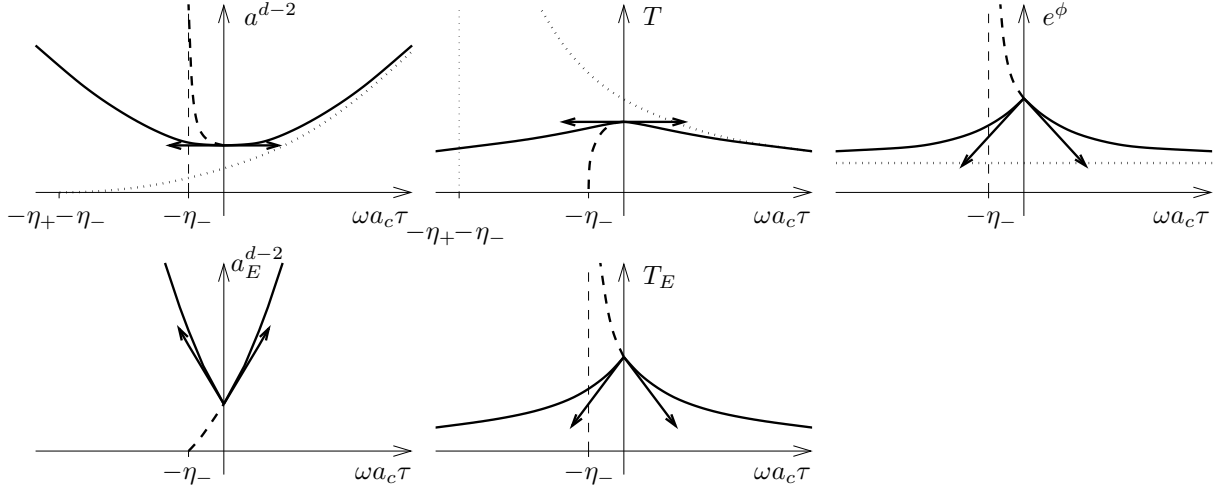
\begin{figure}[t]
%\vspace{.2cm}
\begin{center}
\includegraphics[height=6.3cm]{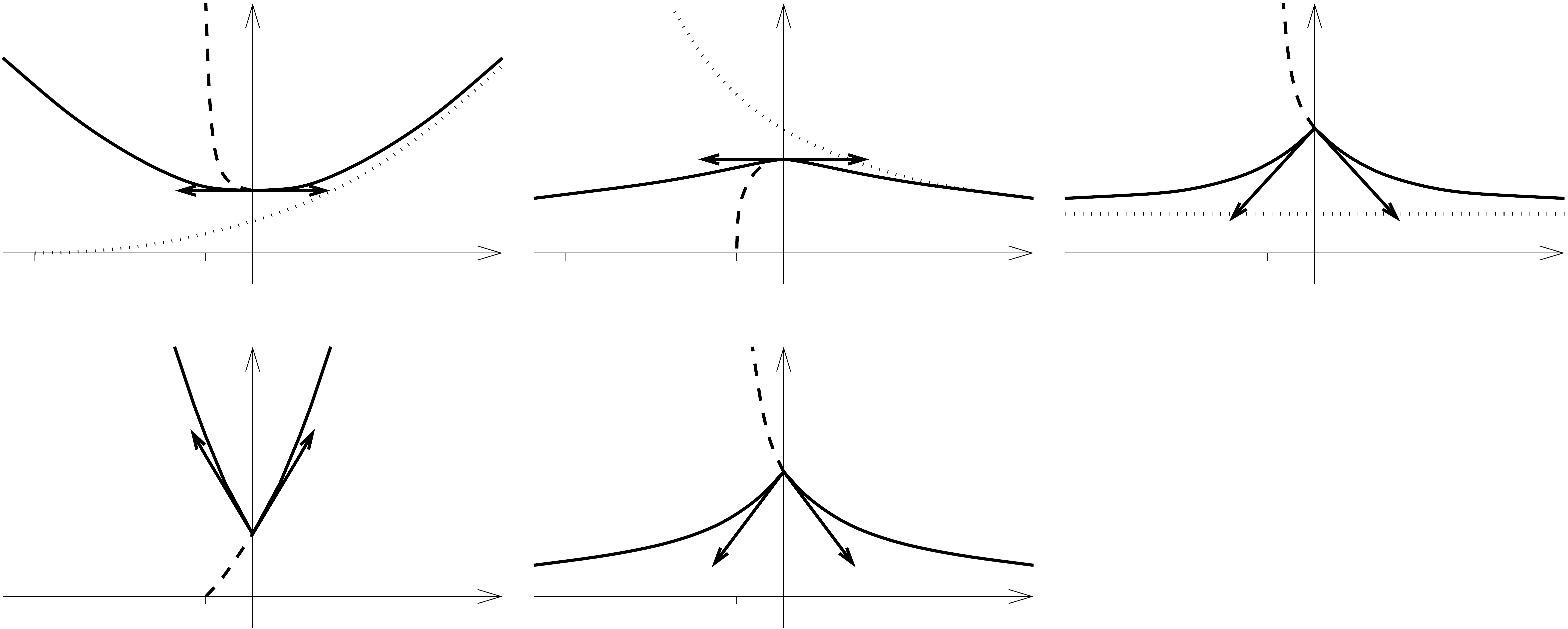}
\end{center}
\begin{picture}(0,0)
\put(91,198){\footnotesize$a^{d-2}$}     \put(242,198){\footnotesize$T$}   \put(393,198){\footnotesize$e^\phi$} 
\put(1,123){\footnotesize$-\eta_+\!\!-\!\eta_-$}     \put(60,123){\footnotesize$-\eta_-$}       \put(132,123){\footnotesize$\omega a_c\tau$}
\put(152,117){\footnotesize$-\eta_+\!\!-\!\eta_-$} \put(212,123){\footnotesize$-\eta_-$}      \put(284,123){\footnotesize$\omega a_c\tau$}
                                                                             \put(364,123){\footnotesize$-\eta_-$}      \put(435,123){\footnotesize$\omega a_c\tau$}
\put(86,105){\footnotesize$a_E^{d-2}$}     \put(242,100){\footnotesize$T_E$}  
\put(60,26){\footnotesize$-\eta_-$}                       \put(133,26){\footnotesize$\omega a_c\tau$}
\put(212,26){\footnotesize$-\eta_-$}                     \put(284,26){\footnotesize$\omega a_c\tau$}
\end{picture}
\vspace{-.8cm}
\caption{\footnotesize \em  For $k=0$ and $d>2$, comparison  between: $(i)$ Classical General Relativity coupled to radiation (doted lines), $(ii)$ Classical General Relativity coupled to radiation and dilaton motion (dashed lines),  and $(iii)$ the superstring picture (solid lines). Case $(i)$ leads to the conventional initial curvature and infinite temperature singularities. In case $(ii)$, the early times of the Universe are out of perturbative control. In case $(iii)$, the winding $\to$ momentum effective field theory phase transition at $\tau=0$ substitutes the previous non-perturbative regime with a  pre-big bang cosmology at weak coupling. In Einstein frame, the whole evolution describes a bounce where the scale factor, the temperature and the dilaton develop conical singularities. Asymptotically, the two phases are radiation dominated.}
%\vspace{0.2cm}
\label{RR}
\end{figure}

For $d>2$, figure \ref{RR}
 makes the comparison between $(i)$ Classical General Relativity coupled to thermal radiation (doted lines), $(ii)$ Classical General Relativity coupled to thermal radiation and non-trivial motion for the dilaton field (dashed lines), and $(iii)$ the thermal  superstring indicating a phase transition at $\tau_c=0$ between the dual effective field theories (solid lines):
\begin{description}
\item $(i)$ In the first case, the scale factor,  
\be
a_{CGR}^{d-2}=a_c^{d-2}\, {\eta_-^{\eta_-}\over \eta_+^{\eta_+}}\, \big(\omega a_c\tau+\eta_++\eta_-\big)^2,
\ee
develops the well known initial singularity at infinite temperature, here at $\omega a_c\tau=-\eta_+-\eta_-$.
\item $(ii)$ When  the system is coupled to a non-trivial dilaton ($\dot\phi\propto e^{2\phi}N/a^{d-1}$), the cosmological evolution takes the form (\ref{sol d}), {\em without} the absolute values in the arguments. In this case, $\sigma(\tau)$ vanishes and bounces at $\tau=0$, so that $\sigma(\tau)\ge 0$ is always satisfied and the whole evolution remains within the framework of the momentum effective field theory. In the Einstein frame, the scale factor and temperature are monotonic and, when we go backward in time, the dilaton drives the Universe into an out of control non-perturbative regime $( \phi\to +\infty$, $a_E\to 0$, $T_E\to +\infty$ when $\omega a_c\tau\to -\eta_-)$. 
\item $(iii)$ The phase transition at $\tau=0$ dictated by string theory effectively substitutes the previous non-perturbative regime in the momentum effective field theory with a perturbative pre-big bang cosmology in  the dual winding effective field theory. In the Einstein frame, the scale factor and temperature satisfy:
\be
a_E^{d-2}=a_c^{d-2}e^{-2\phi_c} \left({\omega a_c\abs\tau\abs\over \eta_+}+1\right)\!\left({\omega a_c\abs\tau\abs\over \eta_-}+1\right), \qquad  T_E={a_cT_c\over a_E}\, .
\ee
\end{description}

In two dimensions, the case $(i)$ of Classical General Relativity  coupled to thermal radiation only does not make sense. In the presence of dilaton motion, the comparison between cases $(ii)$ and $(iii)$ leads to conclusions identical to those in higher dimension. 

%%%%%%%%%%%%%%%%%%%%%%%%%%%%%%%%%%%%%%%%

\subsection{Bouncing cosmology with non-vanishing curvature, $k\ne 0$}
\label{Thk=-1}

We  now turn for $d>2$ to the case of a homogeneous and isotropic Universe with negative spatial curvature $k=-1$. As was the case of $k=0$, the cosmological evolution is compatible with the constraints 1--5 and with localized branes at $\sigma(\tau_c)= 0$ triggering the transition between the winding-like and the momentum-like field theory. The scale factor $a_c$ and dilaton  $\phi_c$ at the critical temperature $T_c$ can be chosen as independent integration constants characterizing the whole solution. 
It is convenient to combine $a_c$ and $\phi_c$ in terms of a parameter $\lambda$ which can be further used to define $\alpha_\pm$,
\be
 \lambda:={2\over d-2}\, n^* \Sigma_d\, e^{2\phi_c}\,a_c^2\,T_c^d \; , \qquad 
\alpha_\pm={\sqrt{d-2}\, \lambda\pm2\sqrt{1+\lambda}\over 2\sqrt{d-1}\sqrt{1+\lambda}+\sqrt{d-2}\, (2+\lambda)},
\ee
satisfying 1$>$$\alpha_+$$>$0 and $\alpha_+$$>$$\alpha_-$$>$$-1$.
These quantities appear explicitly in the algebraic expressions of the scale factor and the dilaton.  
In the conformal gauge (\ref{confor}), $\sigma(\tau)$ and $\phi(\tau)$ take the following form:
\be
\label{soldk=-1}
\begin{array}{ll}
&\dis \sigma={\sign(\tau)\over d-2}\left[ \eta_+\ln \left({e^{{d-2\over 2}\abs\tau\abs}-\alpha_- \, e^{-{d-2\over 2}\abs\tau\abs}\over 1-\alpha_-}\right)-\eta_-\ln\left({e^{{d-2\over 2}\abs\tau\abs}-\alpha_+ \, e^{-{d-2\over 2}\abs\tau\abs}\over 1-\alpha_+}\right) \right],\tesp\\
&\dis \phi=\phi_c+{\sqrt{d-1}\over 2}\left[\ln\left({e^{{d-2\over 2}\abs\tau\abs}-\alpha_- \, e^{-{d-2\over 2}\abs\tau\abs}\over 1-\alpha_-}\right) - \ln \left({e^{{d-2\over 2}\abs\tau\abs}-\alpha_+ \, e^{-{d-2\over 2}\abs\tau\abs}\over 1-\alpha_+}\right)\right] .
\end{array}
\ee
For $\abs\tau\abs\ll 1$, the bounces for the metric and dilaton are respectively smooth and conical: 
\be
\begin{array}{ll}
&\dis \sigma ={\sign(\tau)\,(d-2) \,(2+\lambda)\over 4}\,\, \tau^2+\O(\abs\tau\abs^3),\desp\\
&\dis \phi =\phi_c- {\sqrt{(d-1)(d-2)\,(1+\lambda)}\over 2}~\abs\tau\abs+\O\big(\tau^2).
\end{array}
\ee
The brane tension $\kappa_C$ is given in terms of $a_c$, $\lambda$ and the space-time dimension,
\be
\label{kappa-1}
\kappa_C={2\sqrt{(d-1)(d-2)}\over a_c}\, \sqrt{1+\lambda}.
\ee 
It determines the maximal value of the Ricci scalar, which is obtained at the bounce,
\be
{\cal R}_c={\kappa_C^2 \over 4}=\O(a_c^{-2})+\O(g_c^2).
\ee
The latter is small provided the inverse scale factor and string coupling at the transition  are chosen sufficiently small. In this case, higher derivative terms and higher loop corrections can be consistently neglected throughout the evolution.

At very early and late times ($\abs \tau\abs\gg 1$), the dilaton motion vanishes while the Universe cools and grows to infinity. In the Einstein frame, the scale factor $a_E$, the temperature $T_E$ and the dilaton bounce with conical behavior at the origin $\tau=0$. In the asymptotic limits $\tau\to \pm \infty$, the dilaton motion, the thermal radiation and the spatial curvature contribute to the energy density as $1/a_E^{2(d-1)}$, $1/a_E^d$ and $1/a_E^2$ (in the gauge $N_E=1$). As a result, the cosmological solution describes a bounce between two asymptotically curvature dominated Universes.  As a result, the cosmological solution describes a bounce between two asymptotically curvature dominated Universes.  

\begin{figure}[t]
%\vspace{.2cm}
\begin{center}
\includegraphics[height=6.3cm]{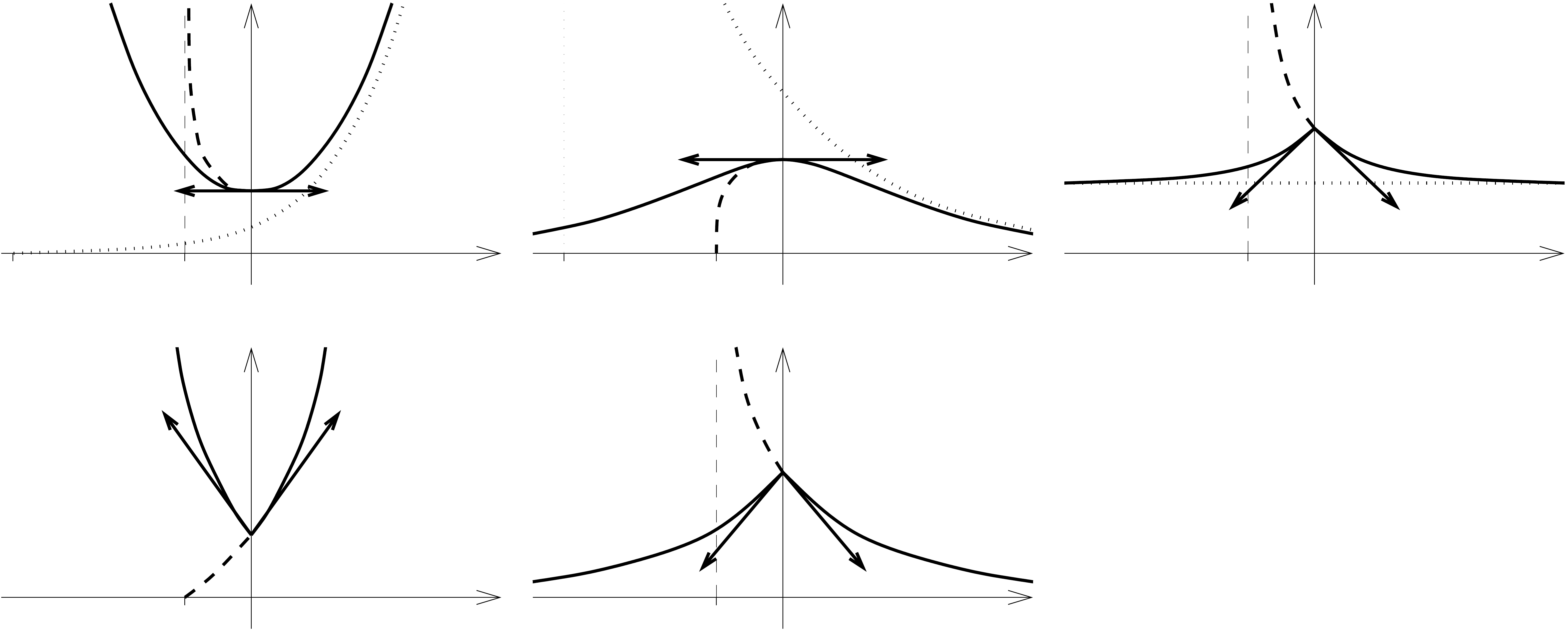}
\end{center}
\begin{picture}(0,0)
\put(91,198){\footnotesize$a^{d-2}$}     \put(242,198){\footnotesize$T$}   \put(393,198){\footnotesize$e^\phi$} 
\put(0,123){\footnotesize$\ln({\alpha_++\alpha_-\over2})$}     \put(54,123){\footnotesize$\ln\alpha_+$}       \put(119,123){\footnotesize$(d-2)\tau$}
\put(151,115){\footnotesize$\ln({\alpha_++\alpha_-\over2})$} \put(206,123){\footnotesize$\ln\alpha_+$}      \put(270,123){\footnotesize$(d-2)\tau$}
                                                                             \put(358,123){\footnotesize$\ln\alpha_+$}      \put(421,123){\footnotesize$(d-2)\tau$}
\put(86,105){\footnotesize$a_E^{d-2}$}     \put(242,100){\footnotesize$T_E$}  
\put(54,26){\footnotesize$\ln\alpha_+$}                       \put(119,26){\footnotesize$(d-2)\tau$}
\put(206,26){\footnotesize$\ln\alpha_+$}                     \put(270,26){\footnotesize$(d-2)\tau$}
\end{picture}
\vspace{-.8cm}
\caption{\footnotesize \em  For $k=-1$ and $d>2$, comparison between: $(i)$ Classical General Relativity coupled to radiation (doted lines), $(ii)$ Classical General Relativity coupled to radiation and dilaton motion (dashed lines),  and $(iii)$ the superstring picture (solid lines). Case $(i)$ leads to the conventional initial curvature and infinite temperature singularities. In case $(ii)$, the early times of the Universe are out of perturbative control. In case $(iii)$, the winding $\to$ momentum effective field theory phase transition at $\tau=0$ substitutes the previous non-perturbative regime with a  pre-big bang cosmology at weak coupling. In Einstein frame, the whole evolution describes a bounce where the scale factor, the temperature and the dilaton develop conical singularities. Asymptotically, the two phases are curvature dominated.}
%\vspace{.2cm}
\label{CC}
\end{figure}
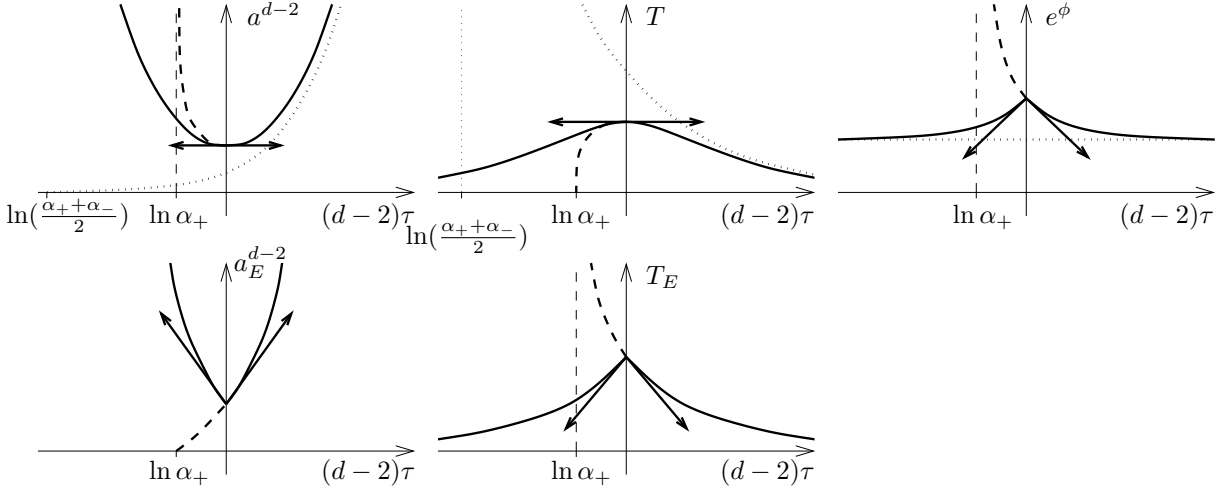

Figure \ref{CC}
makes  the comparison between $(i)$ Classical General Relativity coupled to thermal radiation (doted lines), $(ii)$ Classical General Relativity coupled to thermal radiation and non-trivial motion for the dilaton field (dashed lines), and $(iii)$ the superstring cosmology where two dual effective field theories are connected by a phase transition at $\tau_c=0$ (solid lines). The conclusions are qualitatively similar to those described for  $k=0$:
\begin{description}
\item $(i)$ In the first case, the infinite curvature and temperature singularities occur at the initial time $(d-2)\tau=\ln({\alpha_++\alpha_-\over 2})$,
\be
a_{CGR}^{d-2}=a_c^{d-2}\, {(1-\alpha_+)^{\eta_-}\over (1-\alpha_-)^{\eta_+}}\,\Big(e^{{d-2\over 2}\tau}-{\alpha_++\alpha_-\over 2}\, e^{-{d-2\over 2}\tau}\Big)^2.
\ee
\item $(ii)$ In presence of dilaton motion ($\dot\phi\propto e^{2\phi}N/a^{d-1}$), the cosmological evolution given in (\ref{soldk=-1}) with no absolute values for $\tau$ behaves as follows. $\sigma(\tau)$ vanishes and bounces at $\tau=0$, which implies the evolution is described in the momentum effective field theory only. The dilaton, together with the Einstein frame scale factor and temperature are monotonic. Thus, the early times of the Universe are out of perturbative control, $( \phi\to +\infty$, $a_E\to 0$, $T_E\to +\infty$ when $(d-2)\tau\to \ln\alpha_+)$. 
\item $(iii)$ In the thermal superstring case, the phase transition at $\tau=0$ replaces the above non-perturbative regime with a perturbative pre-big bang cosmology in  the dual winding effective field theory. In the Einstein frame, the scale factor and temperature are 
\be
a_E^{d-2}=a_c^{d-2}e^{-2\phi_c}\left({e^{{d-2\over 2}\abs\tau\abs}-\alpha_+ \, e^{-{d-2\over 2}\abs\tau\abs}\over 1-\alpha_+}\right)\!\left({e^{{d-2\over 2}\abs\tau\abs}-\alpha_- \, e^{-{d-2\over 2}\abs\tau\abs}\over 1-\alpha_-}\right)\!, \;\,  T_E={a_cT_c\over a_E}.
\ee
\end{description}

The case $k=-1$ gives us the possibility to consider the evolution of the Universe at  genus-0, \ie  without  the genus-1 thermal corrections. Indeed, setting $\lambda $=0 in Eq. (\ref{kappa-1}),  the brane tension  at  the tree level approximation  does not vanish,
\be
\kappa^{\rm tr}_C={2\sqrt{(d-1)(d-2)}\over a_c}.
\ee
This fact indicates a non-trivial dilaton motion still exists at this order, provided $d>2$. The explicit solution at this approximation is obtained 
from Eq. (\ref{soldk=-1}) in the limit  $\lambda \to 0$ and can be brought into the form,
 \be
\label{treek=-1}
\begin{array}{ll}
&\dis \sigma={\sign(\tau)\over d-2}\left[ \eta_+\ln \left({\cosh\big({d-2\over 2}\, (\abs \tau\abs +\tau^{\rm tr})\big)\over \cosh\big({d-2\over 2}\,\tau^{\rm tr}\big)}\right)-\eta_-\ln\left({\sinh\big({d-2\over 2}\, (\abs \tau\abs +\tau^{\rm tr})\big)\over \sinh\big({d-2\over 2}\,\tau^{\rm tr}\big)}\right)\right],\tesp\\
&\dis \phi=\phi_c-{\sqrt{d-1}\over 2}\left[ \ln \left({\sinh\big({d-2\over 2}\, (\abs \tau\abs +\tau^{\rm tr})\big)\over \sinh\big({d-2\over 2}\,\tau^{\rm tr}\big)}\right)-\ln\left({\cosh\big({d-2\over 2}\, (\abs \tau\abs +\tau^{\rm tr})\big)\over \cosh\big({d-2\over 2}\,\tau^{\rm tr}\big)}\right) \right] ,
\end{array}
\ee
where we have defined
\be
\tau^{\rm tr}={\ln\left(\sqrt{d-1}+\sqrt{d-2}\right)\over d-2}.
\ee
This evolution  does not describe flat (Minkowski) space, as is the case of  vanishing spatial curvature. Instead, it corresponds to a dual time-dependent curvature dominated cosmology which bounces at $\tau=0$, when the spacelike brane is crossed. The flat (Minkowski) space and the above $k=-1$ tree level cosmology are dual, as follows from the fact that they are both solutions at genus-0 without central charge deficit ($\delta c=0$) of the underlying conformal worldsheet theory.

Given the fact that the notion of spatial curvature $k$ is irrelevant in two dimensions, it is a non-trivial consistency check to recover the two-dimensional solution (\ref{d=2k=0}) by treating the space-time dimension as a continuous parameter and taking the limit $d\to 2$ in (\ref{soldk=-1}). In particular, $\kappa_C\sim \kappa_R$ when $d\to 2$. Since at genus-1 both thermal evolutions with $k=0$ and $k=-1$ tend to the same solution when $d\to 2$, one expects the ``dual-to-flat" classical cosmology (\ref{treek=-1}) to converge to the flat (Minkowski)  genus-0 phase transition in this limit. This is easily checked, which shows that the weak coupling limit $\phi_c\to -\infty$ and the $d\to 2$ limit are commuting.

%%%%%%%%%%%%%%%%%%%%%%%%%%%%%%
%%%%%%%%%%%%%%%%%%%%%%%%%%%%%%%%%
 
\section{Conclusions}
\label{conclu}

The scope of this work was to establish a stringy mechanism able to resolve both  the Hagedorn instabilities of finite temperature superstring theory as well as  the initial curvature singularity of the induced cosmology in arbitrary dimensions. The key ingredients of this mechanism were first isolated in the context of the two-dimensional Hybrid models whose right-moving sector enjoys the \emph{MSDS} structure. The latter ensures boson/fermion degeneracy in the right-moving massive level.

In this paper we have shown that these stringy ingredients are generic in a large class of   $\N_4=(4,0)$ superstring models. Tachyon-free, thermal configurations in $d$-dimensional $\N_4=(4,0)$ models can be constructed in the presence of special ``gravito-magnetic'' fluxes. The fluxes modify the thermal vacuum by injecting into it non-trivial momentum and winding charges, lifting the Hagedorn instabilities of the canonical thermal ensemble. The key property is the restoration of the thermal T-duality symmetry of the stringy thermal system, implying a maximal critical temperature. In all such models there are three characteristic regimes, each with a distinct effective field theory description: Two dual asymptotically cold regimes associated with the light thermal momentum and light thermal winding states, and the intermediate regime where additional massless thermal states appear leading to enhanced Euclidean gauge symmetry. 

Taking into account the genus-0 backgrounds associated with the extra massless states, we have shown that they source Euclidean branes localized at the critical point, which glue the two asymptotic momentum and winding regimes. By utilizing string calculational techniques, we were able to establish that the thermal partition function can be well-approximated by that of massless thermal radiation up to the critical temperature. The partition function exhibits a conical structure as a function of the thermal modulus $\sigma$, generalizing the two dimensional Hybrid result to any dimension. In all regimes, the equation of state is effectively given by
$$
\rho= (d-1) P= (d-1)n^* \Sigma_d \, T_c^d\,  e^{-d\abs \sigma\abs}, 
$$
modulo exponentially suppressed contributions. Both the energy density and pressure are bounded from above attaining their maximal values at the critical temperature. We show explicitly that the conical structure is resolved by the spacelike branes, which  occur when $\sigma=0$. These branes provide a localized negative pressure which is crucial in evading the constraints on realizing singularity-free, bouncing cosmologies, imposed by the singularity theorems of classical general relativity \cite{HE}. Utilizing the above ingredients, we were able to obtain a string effective Lorentzian action covering the three characteristic regimes simultaneously. This action incorporates the spacelike brane that glues together the winding and the momentum regimes. 

Taking into account the localized negative pressure contribution from the branes, as well as the bulk thermal corrections, we have obtained {\it non-singular} analytic cosmological solutions, describing bouncing thermal and curvature dominated Universes. The bounce occurs at the phase transition between the momentum and winding regimes. The cosmological solutions remain perturbative throughout the evolution, provided that the value of the string coupling at the branes is sufficiently small. These bouncing cosmologies are the first higher dimensional examples, where both the Hagedorn singularities as well as the classical Big-Bang singularity are successfully resolved, remaining perturbative throughout the evolution.  

In this work we presented cosmological solutions associated with a single phase transition, with the spacelike branes occurring at a specific point in time. The brane interpretation leads us to believe that consistent cosmological solutions exist associated with a series of phase transitions, described by a distribution of branes localized in several points in time. At each brane the temperature reaches the maximal value.  This work is currently under investigation \cite{FKTT}. 

Having in our disposal exact cosmological solutions, we can explicitly calculate the spectrum of fluctuations at early times, say at the time locations of the branes, determine their propagation at latter cosmological times and compare them to the current and future observational data. This is possible since we have analytical control on the theory describing the brane. This work is currently in progress \cite{BKPPT}.

%%%%%%%%%%%%%%%%%%%%%%%%%%%%%%%%%%%%%
%%%%%%%%%%%%%%%%%%%%%%%%%%%%
  
 \section*{Acknowledgement}
 
 We are grateful to L. Alvarez-Gaume, C. Bachas, R. Brandenberger, D. Luest, S. Patil, J. Troost and especially I. Florakis for fruitful discussions. C.K. and H.P. would like to thank the University of Cyprus for hospitality. N.T. and H.P. acknowledge the Laboratoire de Physique Th\'eorique of Ecole Normale Sup\'erieure for hospitality. N.T. would like to thank the Centre de Physique Th\'eorique of Ecole Polytechnique for hospitality. The work of C.K. and H.P. is partially supported by the ANR 05-BLAN-NT09-573739 and a PEPS contract. The work of C.K., H.P. and N.T. is also supported by the CEFIPRA/IFCPAR 4104-2 project and a PICS France/Cyprus. The work of H.P. is partially supported by the EU contracts PITN GA-2009-237920, ERC-AG-226371 and PICS  France/Greece, France/USA.

%%%%%%%%%%%%%%%%%%%%%%%%%%%%%%%%%%%%%
%%%%%%%%%%%%%%%%%%%%%%%%%%%%
\vspace{.4cm}
%\newpage

\end{document}